\documentclass[prd,%
 reprint,
 amsmath,amssymb,
 aps,
]{revtex4-2}

\usepackage{graphicx}
\usepackage{dcolumn}
\usepackage{bm}

\begin{document}

\preprint{APS/123-QED}

\title{Modeling coupled constellation dynamics for TianQin under self-gravity}


\author{Yuzhou Fang}
\author{Xuefeng Zhang}
 \email{zhangxf38@sysu.edu.cn}
\author{Hongyin Li}
 \email{lihy333@mail.sysu.edu.cn}
\affiliation{MOE Key Laboratory of TianQin Mission, TianQin Research Center for Gravitational Physics and School of Physics and Astronomy, Frontiers Science Center for TianQin, Gravitational Wave Research Center of CNSA, Sun Yat-sen University (Zhuhai Campus), Zhuhai 519082, China. }

\date{\today}

\begin{abstract}
TianQin is a dedicated geocentric mission for space-based gravitational wave (GW) detection. Among its core technologies, the drag-free and pointing control subsystem (DFPCS) - consisting of suspension, drag-free and pointing controls - keeps the two test masses (TMs) centered and aligned within their housings while maintaining drag-free conditions and precise telescope pointing along the laser-arm directions. This results in orbit-attitude coupled dynamics for the constellation. The coupling is made more prominent due to satellite self-gravity, which requires compensation from DFPCS and generally makes the satellites deviate from pure free-fall orbits. Previous studies assumed that the orbit and attitude dynamics could be decoupled in numerical simulation, neglecting the back-action from the closed-loop control to orbit propagation. To address this, we develop a comprehensive model that can propagate the full 9-body (6 TMs + 3 satellites, orbits and attitudes) dynamics inter-dependently under the inter-satellite pointing and drag-free conditions. This paper is threefold. First, we reassess the applicability of the two TMs and telescope pointing scheme to TianQin using the new model, and confirm the previous conclusion. Second, to meet the constellation stability requirements, it is found that the DC common self-gravity in the flight direction should be minimized, or kept close for the three satellites. Finally, we simulate the long-range light path between two TMs/satellites with a precision of sub-pm/Hz$^{1/2}$, and the results support the decoupling of the closed-loop dynamics and high-precision orbit for computational efficiency. The method is instrumental to other future missions where the orbit-attitude coupling needs careful consideration. 
\end{abstract}

\maketitle

\section {Introduction}\label{sec: intro}
The current TianQin design assumes high Earth orbits with the 1.7 $\times$ 10$^5$ km baselines between three satellites \cite{Luo2016}. It employs a constellation consisting of six reference objects in geodesics, specifically the TMs in the state of free fall along the laser arms. The nearly equilateral triangle constellation stands almost vertically to the ecliptic \cite{Tan2020}. High-precision laser interferometry tracks distance changes between the well-protected TMs in separate drag-free controlled satellites, within a preliminary detection frequency range from 10$^{-4}$ Hz to 1 Hz.

\begin{figure}[ht]
    \centering
    \includegraphics[width=0.48\textwidth]{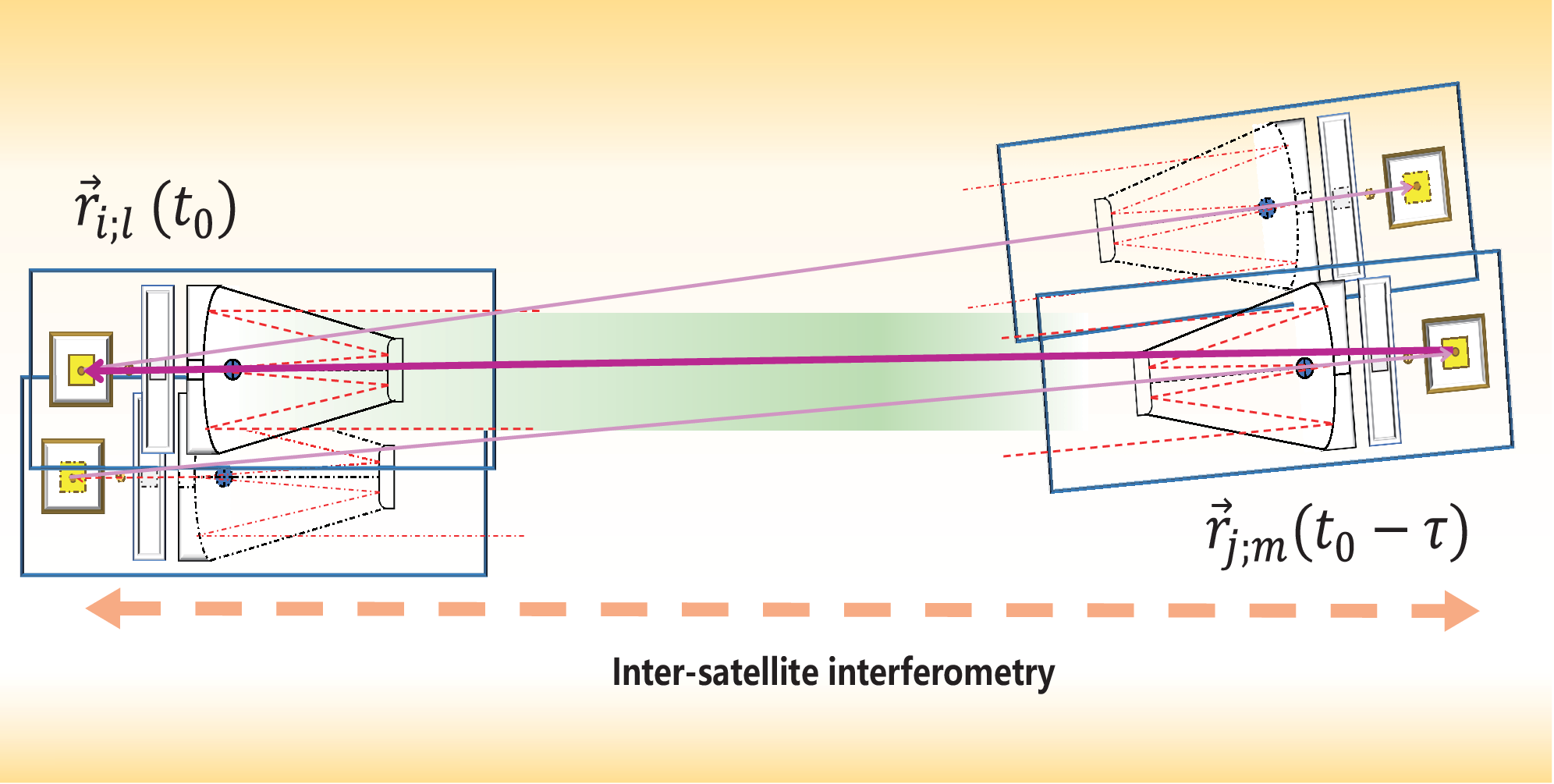}
    \caption{Illustrative diagram of the long-range light path in split measurement. The satellite attitude at the current global time $t_0$ is determined by the light signal transmitted from the remote satellite at the previous time $t_0 - \tau$, where $\tau$ represents the light travel time between two distant TMs' nominal positions (i.e., the housing center). The propagation path of the light through space is depicted by the dark purple vector. }
    \label{fig:threePartMeasure}
\end{figure}

During the scientific operations of space-based GW detection, the control system should be considered to minimize control actuations on the end mirrors of each arm \cite{LISA2017, armano2019lisaDFAC}. After that, the interferometric measurement of the distance changes between pairs of free-falling TMs is broken into three different measurements \cite{LISA2024}. That is, TM–to–satellite (local), satellite-to-satellite (long-range light path, see Fig. \ref{fig:threePartMeasure}) and satellite–to–TM (distant). The requirement to break down the measurements causes imperfections, mainly as a result of the optical misalignments and the residual noises of satellite control (also known as the attitude jitters). These misalignments can result in significant cross-couplings, called tilt-to-length (TTL) couplings \cite{chwalla2020optical}, which restrict the instrument's sensitivity within the millihertz band, unless hardware or post-processing adjustments are made \cite{paczkowski2022postprocessing}. Therefore, gaining an in-depth understanding of the system dynamics is crucial. To achieve this, we require a comprehensive simulation of the closed-loop dynamics along with the high-precision orbit propagation. These simulations, including the dynamics of the satellite and the TMs, can be integrated into the interferometer data streams to facilitate a more comprehensive simulation to study and suppress instrument noises \cite{heisenberg2023lisa, wanner2024depth, dam2022simulations}.

\begin{figure}[ht]
    \centering
    \includegraphics[width=0.48\textwidth]{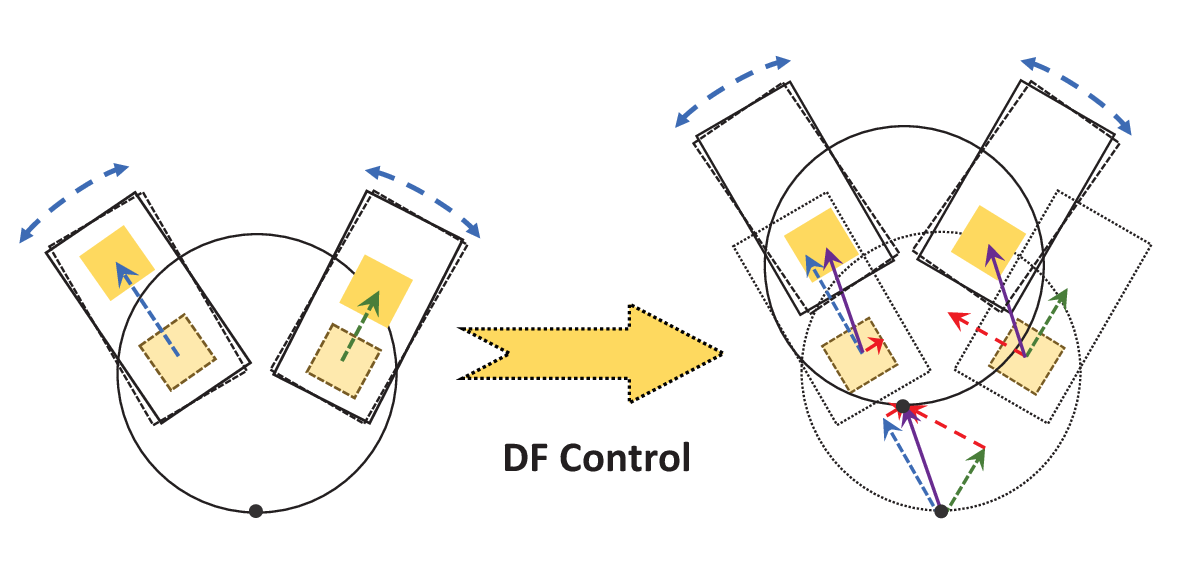}
    \caption{Schematic description of the drag-free control designed to enable free-fall of the two TMs along the sensitive axes in each satellite. The red vectors indicate the electrostatic controls applied to the TMs, allowing the satellite to simultaneously follow both of them as the purple vector indicated. }
    \label{fig:dragFreedemo}
\end{figure}

In the context of TianQin, the control of pointing includes both the attitude of the satellites and the alignment of the two moving optical subassemblies (MOSAs). This control restricts the orientation of the satellite relative to the incoming laser wave fronts emitted by distant satellites, thereby securing the integrity of the triangular constellation dynamics. For details, as input for each satellite, this control takes in differential wavefront sensing (DWS) data, then modifies the satellite orientation with respect to the constellation (3 degrees of freedom) and compensates for stray torques with the micropropulsion system. In addition, for proper alignment, the two MOSA axes will symmetrically rotate to compensate for the constellation breathing angle (1 degree of freedom). For the TMs, in order to keep them centered and aligned within the electrode housings, the controls along the axes defined by the optical links are maintained solely through satellite actuations. The electrostatic suspension controls actively manage the remaining degrees of freedom. This approach results in a significant reduction in the suspension control forces and torques applied on the nonsensitive axes. Subsequently, it mitigates the effect of actuation crosstalk in the sensitive axes, which would lead to an important contribution to the acceleration noise budget \cite{inchauspe2022new}. One potential drawback is that the orbits of satellites are influenced not only by gravitational forces but also by the unique effects of the satellites' self-gravity and pointings, as demonstrated in Fig. \ref{fig:dragFreedemo}. As a consequence of the implemented control strategies, the entire constellation dynamics, comprising the three satellites, six MOSAs and six TMs, are inherently coupled.

Unlike heliocentric orbit missions, such as LISA, geocentric missions experience attitude propagation under the condition of a relatively fast constellation evolution. It is important to simulate and study the attitude and orbit dynamics, to evaluate if this dynamic propagation might impact the detection frequency band and also the control magnitude under self-gravity. In geocentric orbit, the perturbations induced by celestial bodies and the inertial acceleration are counterbalanced by electrostatic and microthrust accelerations at a magnitude of 10$^{-13}$ m/s$^2$ \cite{fang2024payload}. However, self-gravity, identified as a kind of DC acceleration acting on the TMs, is expected to be on the magnitude of 10$^{-9}$ m/s$^2$ \cite{armano2016constraints}. This represents a significantly higher magnitude that poses challenges for the control system to adequately compensate. For the suspension part, the self-gravity effect on the control allocation for the nonorthogonal sensitive axes architecture is yet not fully studied under the coupled constellation dynamics. For the drag-free part, the effect on the satellite's orbit has been briefly examined in \cite{martens2021trajectory, joffre2021lisa}. This includes the influence of a 10$^{-9}$ m/s$^2$ self-gravity on the entire satellite. However, it overlooks the satellite architecture and the coupled constellation dynamics. 

Before the mission begins, all procedures must be examined during the conceptual study. This examination includes precise orbital computations, assessments of constellation stability, integrated simulations, generation of raw data, and pre-processing. All of these steps are need to be validated through thorough modeling and analysis. For the design of the TianQin mission, we already have a refined series of studies on geocentric orbits and environment \cite{ye2021eclipse, zhang2021effect, luo2022effect, jiao2023effect, jing2022plasma, liu2024solar}. For the dynamics perspective, previous research has treated orbit and attitude simulations with some decoupling assumption \cite{ye2023novel, zhang2024nonlinear}, without in-depth consideration of their coupling effects. This may lead to some round-off errors in simulation, particularly in the context of a high-precision interferometric measurement mission. To better understand the dynamic properties of the TianQin satellites, the study of the coupled dynamics of the constellation is represented in this paper. With the method based on numerical orbits and the coupled constellation dynamics derived from the pointing scheme. 

This paper aims to evaluate the coupled constellation dynamics in detail, which can be divided into the following structures. In Sec. \ref{sec: setup}, we list all the input of our simulation. After that, we present a comprehensive coupled dynamics model in section \ref{sec: orbit and attitude coupled}. The implementation of the coupled dynamics for numerical simulation is elaborated thoroughly in Sec. \ref{sec: implementation}. Finally, Sec. \ref{sec: selfGraDYN} presents the three main conclusions that can be derived from the simulation, i.e., a) the reassessment of the two TMs and telescope pointing scheme in geocentric orbits, b) the self-gravity design suggestions, and c) the decoupling argument between closed-loop dynamics and high-precision orbit propagation.


\section{Model setup} \label{sec: setup}
The assessment of TianQin's dynamics needs detailed models of celestial bodies and the satellite architecture. For the celestial models part, it is imperative that the force models employed are both comprehensive and timely, incorporating all relevant gravitational disturbances, especially those perturbations that may fall within the detection band. Thus, in this study, we inherit the refined force models and the optimized initial orbital elements in \cite{zhang2021effect}, which are summarized in Table \ref{tab:orbit} and \ref{tab:models}. 
\begin{table}[ht]
\caption{\label{tab:orbit} This research adopted the optimized initial orbital elements of the TianQin constellation in the J2000-based Earth-centered equatorial coordinate system at the epoch 06 Jun. 2004, 00:00:00 UTC for evaluation purposes. Here, $a$ denotes the semi-major axis, $e$ the eccentricity, $i$ the inclination, $\Omega$ the longitude of the ascending node, $\omega$ the periapsis argument and $\nu$ the truly anomaly \cite{zhang2021effect}.}
\begin{ruledtabular}
\begin{tabular}{ccccccc}
  & $a$ & $e$ & $i$ & $\Omega$ & $\omega$ & $\nu^{\rm ini}$ \\ 
\hline
Sat1 & $100000.0$ km & 0 & $74.5^\circ$ & $211.6^\circ$ & $0^\circ$ & $ 30^\circ$ \\
Sat2 & $100009.5$ km & 0 & $74.5^\circ$ & $211.6^\circ$ & $0^\circ$ & $150^\circ$ \\
Sat3 & $ 99995.0$ km & 0 & $74.5^\circ$ & $211.6^\circ$ & $0^\circ$ & $270^\circ$ \\
\end{tabular}
\end{ruledtabular}
\end{table}

\begin{table}[ht]
\caption{\label{tab:models}The list of force models implemented in the simulation \cite{zhang2021effect}. }
\begin{ruledtabular}
\begin{tabular}{lc}
Models                         & Specifications \\
\hline
Solar system ephemeris         & JPL DE430\\
Earth's precession \& nutation & IAU 2006/2000A\\
Earth's polar motion           & EOP 14 C04\\
Earth's static gravity field   & EGM2008 ($n=12$)\\
Solid Earth tides              & IERS (2010)\\
Ocean tides                    & FES2004 ($n=10$)\\
Solid Earth pole tide          & IERS (2010)\\
Ocean pole tide                & Desai (2003)\\
Atmospheric tides              & Biancale \& Bode (2003)\\ 
Moon's libration               & JPL DE430\\
Moon's static gravity field    & GL0660B ($n=7$)\\
Sun's orientation              & IAU \\
Sun's $J_2$                    & IAU \\
relativistic effect            & post-Newtonian\\
\end{tabular}
\end{ruledtabular}
\end{table}

\begin{figure}[ht]
    \centering
    \includegraphics[width=0.48\textwidth]{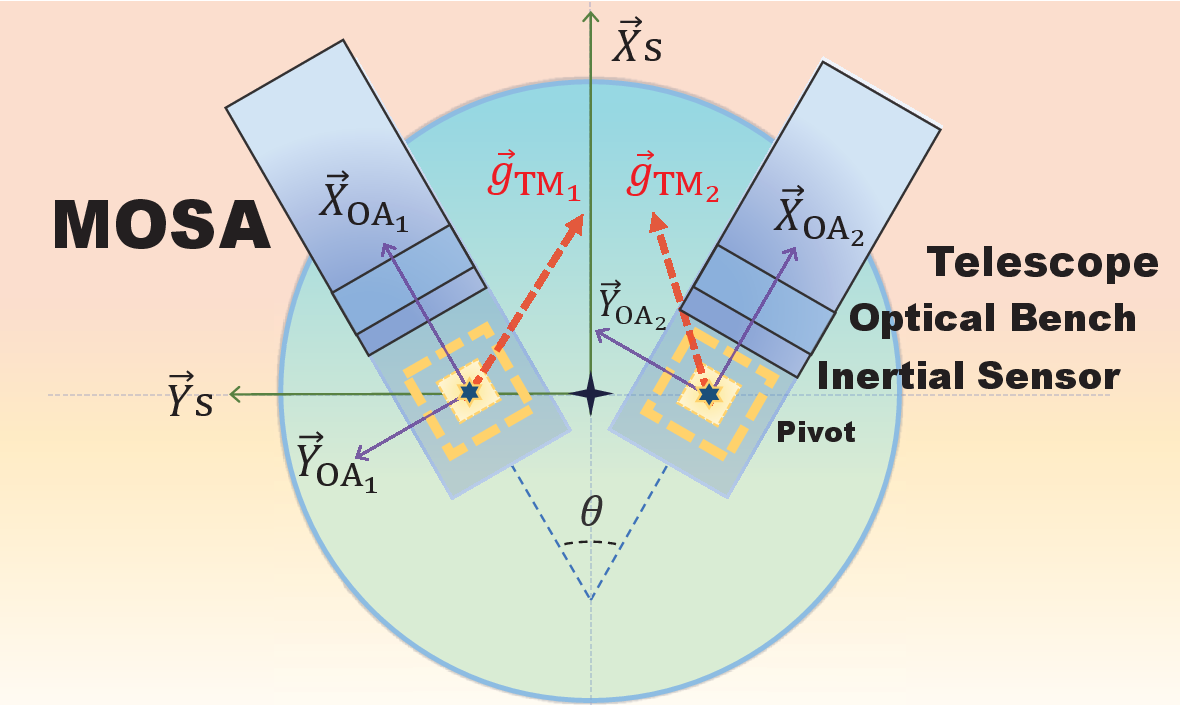}
    \caption{Demonstration diagram of the satellite architecture. The red vectors $\vec{g}_{TM_1}$ and $\vec{g}_{TM_2}$ are the differential accelerations of each TM relative to the satellite (including celestial gravity acceleration, inertial acceleration, and DC self-gravity acceleration). }
    \label{fig:satDemo}
\end{figure}

The satellite architecture encompasses the arrangement of the TMs as well as the location of the MOSA pivots. In this setting, we adopt the previously fine-tuned mechanical parameters (one can find in Ref. \cite{fang2024payload}). It arranges the centers of mass (CoMs) of the two TMs to be collinear and equidistant from the satellite CoM (separating them by approximately 40 cm) and aligns the pivots of the assembly with the centers of the electrode housing (EH), as specified in Table \ref{tab:SCparam} and demonstrated in Fig. \ref{fig:satDemo}. With this arrangement of geometric parameters, the nominal electrostatic control accelerations on the TMs are diminished, as well as the compensated accelerations of drag-free actuations on the satellite.
\begin{table}[htb]
\caption{\label{tab:SCparam} Geometric parameters used in the simulations}
\begin{ruledtabular}
\begin{tabular}{ccc}
Symbols & Parameters & Values
\\ 
\hline
$d_\mathrm{TM}$ & \text{Distance of two TMs}         & $40$ cm\\
$d_\mathrm{p}$  & \text{Distance of pivots from EHs} & $ 0$ cm\\
\end{tabular}
\end{ruledtabular}
\end{table}

\section{Coupled constellation dynamics}\label{sec: orbit and attitude coupled}
The orientations of satellites will meet stringent requirements as a result of the implementation of pointing control. Consequently, the attitude of an individual satellite is adjusted along with the whole constellation. Meanwhile, drag-free control affects the orbits of the satellites, especially under self-gravity, thereby impacting the constellation. The dynamics of this coupled constellation can be mathematically represented by examining both the attitude and orbit components. This following section will introduce this coupled dynamics in detail.

\subsection{Attitude dynamics} \label{sec:OAdependence1}
In order to describe the dynamics process of the three satellites, six MOSAs, and six TMs. We first briefly introduce the coordinates that have been used in this paper. They are the target coordinates of the objects determined by their mathematical relationship with the orbits. They are defined as follows:
\begin{itemize}
   \item The $\mathcal{S}_i$ frame ($i=1,2,3$) describes the target/nominal attitude for one satellite;
   \begin{itemize}
       \item The origin coincides with the nominal orbit of the satellite.
       \item $\vec X_{S_i}$ is the bisector of two incoming laser beams. 
       \item $\vec Z_{S_i}$ is orthogonal to the constellation plane.
       \item $\vec Y_{S_i}$ is built from the cross product of the two above.
   \end{itemize}
   
   \item The $\mathcal{OA}_l$ frame ($l=1,2$) describes the target attitude of one optical assembly. They also provide the reference coordinates for verifying the position of the TMs when they are in the nominal states.
   \begin{itemize}
       \item The origin is at the center of the electrode housing.
       \item $\vec{X}_{\mathrm{OA}_l}$ points at the distant satellite (incoming beam directions). It can be deduced from a rotation of the unit vector $\vec X_{S_i}$ around $\vec Z_{S_i}$ of an angle equal to half the angle of incoming laser beams.
       \item $\vec{Z}_{\mathrm{OA}_l}$ aligns with $\vec{Z}_\mathrm{S}$.
       \item $\vec{Y}_{\mathrm{OA}_l}$ is determined by the right-hand rule.
   \end{itemize}
\end{itemize}

\begin{figure}[ht]
    \centering
    \includegraphics[width=0.48\textwidth]{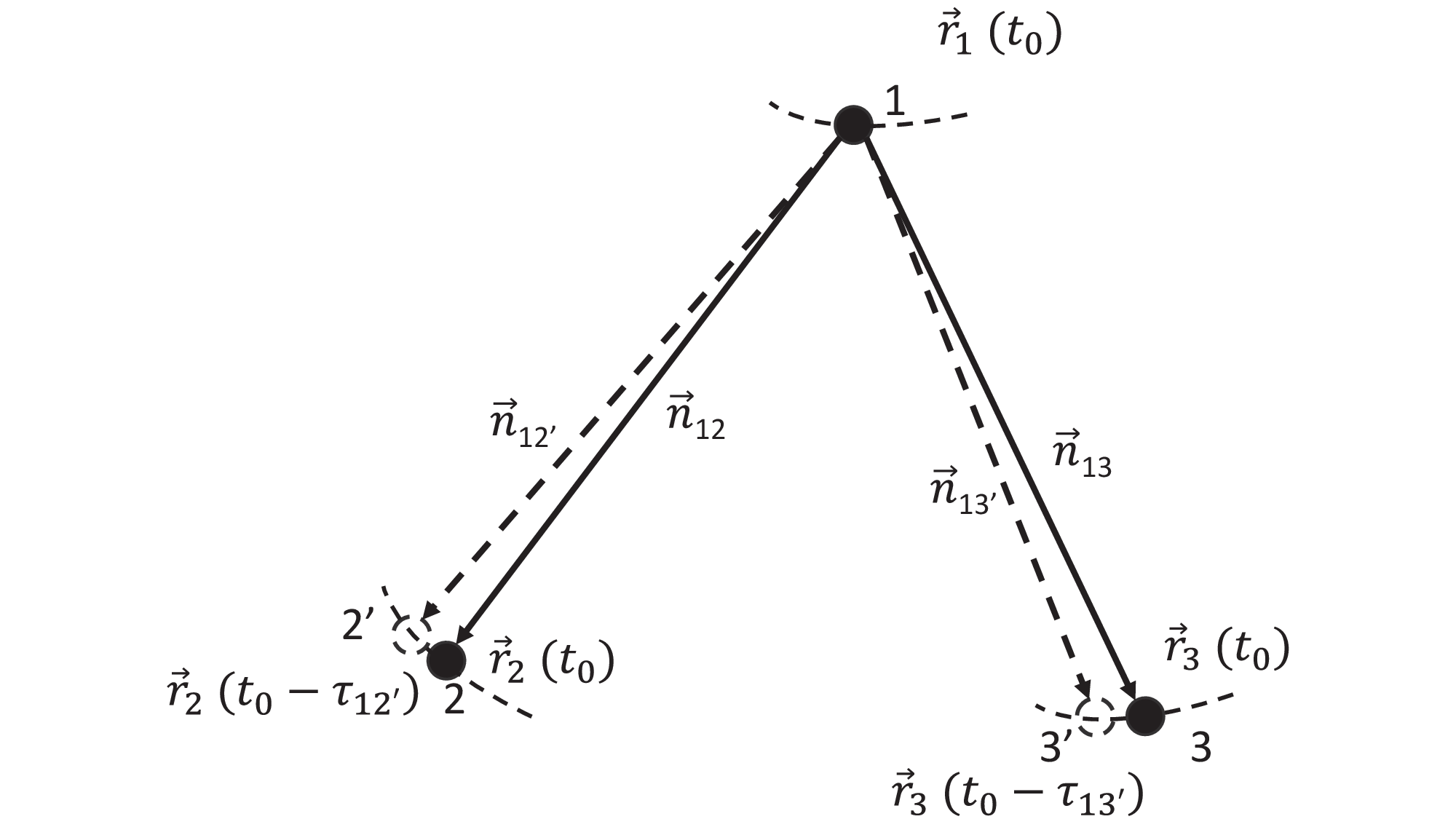}
    \caption{Sketch of the attitude determination with global (solid line) and local (dashed line) constellation plane. The black dots represent the real-time satellite position and the blank dots represent the last-time satellite position where they transmitted the light to the current local satellite.}
    \label{fig:orbitAttitudeCoupledemo}
\end{figure}

The nominal attitude of each satellite is determined by the orientations of the two incoming laser beams, which are related to the constellation (i.e. the position of two laser emitted satellites). It is vital to account for the pointing effects caused by the light traveling time (LTT) in the determination of the nominal attitudes. In Fig. \ref{fig:orbitAttitudeCoupledemo}, $\vec{n}_{ij}$ and $\vec{n}_{ik}$ (where $i,j,k$ = 1,2,3) are the vectors that represent the directions of the instantaneous orbit positions from Sat$_i$ to Sat$_j$ and Sat$_k$. The $\vec{n}_{ij'}$, $\vec{n}_{ik'}$ are the vectors that represent the directions from Sat$_i$ to the laser transmitted time position of Sat$_{j'}$ and Sat$_{k'}$. The LTTs, denoted by $\tau_{ij'}$ and $\tau_{ik'}$, read
\begin{equation} \label{eq:lightTravelTime}
\begin{aligned}
    \tau_{ij'}={|{{\vec{r}_{j}(t_0-\tau_{ij'})}-{\vec{r}_{i}(t_0)}}|}/c,\\
    \tau_{ik'}={|{{\vec{r}_{k}(t_0-\tau_{ik'})}-{\vec{r}_{i}(t_0)}}|}/c,
\end{aligned}
\end{equation}
where $t_0$ indicates an arbitrary time stamp. The pointing unit vectors $\vec{n}_{ij}$, $\vec{n}_{ik}$, $\vec{n}_{ij'}$, and $\vec{n}_{ik'}$ can be defined by
\begin{equation}\label{eq:geometrypointing}
\begin{aligned}
    \vec{n}_{ij} = \frac{\vec{r}_j(t_0) - \vec{r}_i(t_0)}{|\vec{{r}_j}(t_0) - \vec{{r}_i}(t_0)|},\\
    \vec{n}_{ik} = \frac{\vec{r}_k(t_0) - \vec{r}_i(t_0)}{|\vec{{r}_k}(t_0) - \vec{{r}_i}(t_0)|},
\end{aligned}
\end{equation}
and 
\begin{equation}\label{eq:realpointing}
\begin{aligned}
    \vec{n}_{ij'} = \frac{\vec{r}_j(t_0-\tau_{ij'}) - \vec{r}_i(t_0)}{c\tau_{ij'}},\\
    \vec{n}_{ik'} = \frac{\vec{r}_k(t_0-\tau_{ik'}) - \vec{r}_i(t_0)}{c\tau_{ik'}}.
\end{aligned}
\end{equation}

The unit vectors specified by Eqs. (\ref{eq:geometrypointing}) are viewed within the frame of the global constellation plane. In this framework, the current orientations of the satellites are determined by the instantaneous inter-satellite connection, while the angular velocity of the satellites' motion and the MOSAs' rotational speed are closely locked to the constellation propagation. The mathematical relationship is derived as follows. First, the position of the incenter of the global constellation can be represented by the orbital positions of the three satellites ${\vec{R}_{i/j/k}}$,
\begin{equation}
    \vec{r}_{\mathrm{inc}} = \frac{{{r}_{23}}{\vec{R}_1}+{{r}_{31}}{\vec{R}_2}+{{r}_{12}}{\vec{R}_3}}{r_{12}+r_{23}+r_{31}}, 
\end{equation}
with
\begin{equation}\label{eq:OAinput1}
    \vec{r}_{ij}={\vec{R}_j}-{\vec{R}_i},\quad r_{ij}=|\vec{r}_{ij}|.
\end{equation}
Since the global constellation only considers the satellite positions at the same time, we ignore $t_0$ in the expression. The breathing angle of this constellation can be determined by
\begin{equation} \label{eq:breathingAngle}
    \alpha_i=\arccos(\vec n_{ij} \cdot \vec n_{ik}).
\end{equation}

Now, the nominal attitudes of the satellites are represented by the coordinate $S_i$, where $\vec{X}_{S_i}$ can be defined by
\begin{equation}
    \vec X_{\mathrm{S}_i} \equiv \frac{\vec{r}_{\mathrm{inc}}-\vec{R}_{i}}{|\vec{r}_{\mathrm{inc}}-\vec{R}_{i}|}, 
\end{equation}
and $\vec Z_{S_i}$ are given by
\begin{equation}
    \vec Z_{\mathrm{S}_i} \equiv \frac{\vec n_{ij} \times \vec n_{ik}}{\sin\alpha_i},
\end{equation}
where $[i, j, k]$ must be an even permutation of $[1, 2, 3]$. The nominal orientations of the MOSAs and TMs can therefore be determined on the basis of the defined coordinates. The attitude dynamics, such as angular velocity and acceleration, can also be defined by the three satellites' velocities and accelerations. One can see in the Appendix. \ref{sec:AttitudeDYN} or \cite{heisenberg2023lisa, inchauspe2015lisa, fang2024payload} for more details. 

The unit vectors, as defined by Eqs. (\ref{eq:realpointing}), are conducted within the framework of the local constellation plane. This computation demands significantly more resources compared to the global constellation approach, but is more realistic. We will give a more detailed discussion of using this equation in attitude-orbit coupled dynamics in Sec. \ref{sec:decoupledComputation}.

\subsection{Orbit propagation} \label{sec:OAdependence2}
To achieve the drag-free state of the two nonorthogonal sensitive axes, the CoM of the satellite will track the accelerations of the two TMs after suspension control. The following part details how the attitude of the satellite affects its orbit. 

First, the orbital dynamics of the satellite $i$ in the inertial frame is given by
\begin{equation} \label{eq:satTrans}
\ddot{\vec{r}}_i = \vec{a}_i(\vec{r}_i,\dot{\vec{r}}_i) + \vec{a}_{c,i} + \vec{a}_{dis,i},
\end{equation}
where $\vec{a}_i(\vec{r}_i,\dot{\vec{r}}_i)$ denotes the gravity acceleration of the satellite. Given that the additional controls on the satellite are initially activated by the dynamics of the TMs. Thus, the calculation of the control accelerations on the satellite should begin with the nominal suspension control of the TMs $^{OA_l}{\vec{f}}_{\mathrm{c},l}$, which can be written as (for a detailed deduction, see \cite{fang2024payload})
\begin{widetext}
\begin{equation}\label{eq:TMforce}
\begin{split}
\frac{^{OA_l}{\vec{f}}_{\mathrm{c},l}}{m_{\mathrm{TM}_{l}}} & = \frac{^{OA_l}{\vec{f}}_{\mathrm{c,S}}}{m_\mathrm{S}} - T^{OA_l}_{S}\{({^{S}\vec{a}_{\mathrm{TM}_{l}}}-{^{S}\vec{a}_{\mathrm{S}}}) -2 \; {^{S}\vec{\omega}_\mathrm{S}}\times[{^{S}\vec{\omega}_{\mathrm{OA}_l}}\times({^{S}{\vec{r}}_{0,l}}-{^{S}{\vec{r}}_{\mathrm{p},l}})] -{{^{S}\dot{\vec{\omega}}_{\mathrm{OA}_l}}\times({^{S}{\vec{r}}_{0,l}}-{^{S}{\vec{r}}_{\mathrm{p},l}})}
\\
& -{^{S}\vec{\omega}_{\mathrm{OA}_l}}\times[{^{S}\vec{\omega}_{\mathrm{OA}_l}}\times({^{S}{\vec{r}}_{0,l}}-{^{S}{\vec{r}}_{\mathrm{p},l}})] -{{^{S}\vec{\omega}_\mathrm{S}}\times({^{S}\vec{\omega}_\mathrm{S}}\times{^{S}{\vec{r}}_{0,l}})} - {^{S}\dot{\vec{\omega}}_\mathrm{S}}\times{^{S}{\vec{r}}_{0,l}}\}.
\end{split}
\end{equation}
\end{widetext}
The terms ${^{S}\vec{a}_{\mathrm{TM}_{l}}}$ and ${^{S}\vec{a}_{S}}$ are the gravitational pull on $\mathrm{TM}_l$ and the satellite in the corresponding $\mathcal{S}$ frame. ${^{OA_l}\vec{f}_\mathrm{c,S}}$ is the control forces on the satellite in the $\mathcal{OA}_l$ frame. The symbol $T^{OA_l}_{S}$ is the transfer matrix from the $\mathcal{S}$ frame to the $\mathcal{OA}_l$ frame. The terms ${^{S}\vec{\omega}_\mathrm{S}}$ and ${^{S}\dot{\vec{\omega}}_\mathrm{S}}$ denote the angular velocity and angular acceleration of the satellite. Similarly, the terms ${^{S}\vec{\omega}_{\mathrm{OA}_l}}$ and ${^{S}\dot{\vec{\omega}}_{\mathrm{OA}_l}}$ denote the angular motions of MOSA$_l$. It is noticed that all the attitude dynamics terms of the satellite and MOSAs, transfer matrix and gravity accelerations can be calculated from the orbit information (see Appendix. \ref{sec:AttitudeDYN}). The terms TM ${^{S}{\vec{r}}_{\mathrm{p},l}}$, $^{S}{\vec{r}}_{0,l}$ are the pivot position and the EH center position with respect to the satellite frame. The Eq. (\ref{eq:TMforce}) can be rewritten as
\begin{equation}\label{eq:TMforcesimple}
 {T_S^{OA_l}} \: {^{S}\!{\vec{A}}_{\mathrm{c,S}}} = {^{OA_l}{\vec{A}}_{\mathrm{c},l}} + T^{OA_l}_{S} \: {^{S}\vec{g}_{l}},
\end{equation}
here ${^{OA_l}{\vec{A}}_{\mathrm{c},l}}$ is the suspension control acceleration on TM$_l$, and ${^{S}\!{\vec{A}}_{\mathrm{c,S}}}$ is the control acceleration that the satellite needs to follow. Moreover, ${^{S}\vec{g}_l}$ is the TM acceleration in the $\mathcal{S}$ frame given by the terms in the curly brackets of Eq. (\ref{eq:TMforce}).  

The suspension controls of two TMs primarily concentrate on the differential component. This process can be described by the following steps. The differential acceleration for two TMs $\Delta^{S}\vec{g}$ described in the satellite frame is defined by
\begin{equation}\label{eq:E}
    {\Delta {^{S}\vec{g}}} = {^{S}\vec{g}_{2}}-{^{S}\vec{g}_{1}}.
\end{equation}
Moreover, we use $^S\vec{G}_l$ to denote the accelerations of the two TMs after the suspension controls in the $\mathcal{S}$ frame, which read as
\begin{equation}\label{eq:GE}
\begin{aligned}
    ^S\vec{G}_1 = {^{S}\!\vec{A}_{\mathrm{c},1}} + {^{S}\!\vec{g}_{1}},\\
    ^S\vec{G}_2 = {^{S}\!\vec{A}_{\mathrm{c},2}} + {^{S}\!\vec{g}_{2}}.
\end{aligned}
\end{equation}
There are three conditions to be satisfied by the nominal DFPCS, i.e., a) both TMs having equal accelerations in the $\mathcal{S}$ frame (i.e., $^S\vec{G}_1$ = $^S\vec{G}_2$), b) no suspension implemented along the sensitive axes $\vec{X}_{\mathrm{OA}_l}$, and c) the control forces along the $z$-axes of the two TMs being equal with opposite signs to minimize the differential control actuations on the two TMs. Therefore, we can obtain the required electrostatic acceleration on $\mathrm{TM}_l$ in the $\mathcal{OA}_l$ frame as
\begin{equation}\label{eq:E1}
   {^{OA_1}{\vec{A}}_{\mathrm{c},1}}=\frac{\Delta{^{S}\!g_\mathrm{y}}\sin\frac{\theta}{2}-\Delta{^{S}\!g_\mathrm{x}}\cos\frac{\theta}{2}}{\sin\theta}{\vec{Y}_{\mathrm{OA}_1}} + \frac{\Delta{^{S}\!g_\mathrm{z}}}{2}{\vec{Z}_{\mathrm{OA}_1}},
\end{equation}
\begin{equation}\label{eq:E2}
    {^{OA_2}{\vec{A}}_{\mathrm{c},2}}=\frac{-\Delta{^{S}\!g_\mathrm{y}}\sin\frac{\theta}{2} - \Delta{^{S}\!g_\mathrm{x}}\cos\frac{\theta}{2}}{\sin\theta}{\vec{Y}_{\mathrm{OA}_2}} - \frac{\Delta {^{S}\!g_\mathrm{z}}}{2}{\vec{Z}_{\mathrm{OA}_2}},
\end{equation}
where $\theta$ are the angle of the two telescope pointing, see Fig. \ref{fig:satDemo} (we assume the two MOSA rotate symmetrically). The presentation of the nominal actuation acceleration on the satellite reads
\begin{equation}\label{eq:Greal}
\begin{aligned}
    ^S\vec{G} & = \frac{ {^{S}\!{g}_{\mathrm{x},2}} + {^{S}\!{g}_{\mathrm{x},1}}-{\tan\theta} \; {\Delta{^{S}\!g_\mathrm{y}}}}{2}\vec{X}_\mathrm{S}
    \\
    & +\frac{ {^{S}\!{g}_{\mathrm{y},2}} + {^{S}\!{g}_{\mathrm{y},1}} - {\cot\theta} \; {\Delta{^{S}\!g_\mathrm{x}}}}{2}\vec{Y}_\mathrm{S} + \frac{g_{\mathrm{z},2}+g_{\mathrm{z},1}}{2}\vec{Z}_\mathrm{S}.
\end{aligned}
\end{equation}

Since the satellite will track the common acceleration of the two TMs, and the nonconservative disturbances on the satellite are compensated (i.e., drag-free control). Therefore, the satellite acceleration under nominal control of Eq. (\ref{eq:satTrans}) can be expressed as 
\begin{equation}\label{eq:coreEquation}
\ddot{\vec{r}}_i =  \vec{a}_i(\vec{r}_i,\dot{\vec{r}}_i) + \vec{G}_i,
\end{equation}
where $\vec{G}_i$ is the acceleration we derived in the Eqs. (\ref{eq:Greal}) expressed in the inertial frame. 

To sum up, the Eq. (\ref{eq:coreEquation}) represents the accelerations of the satellites after drag-free control, which are determined not only by the local gravitational acceleration (the first term on the right hand side) but also by the attitude dynamics, satellite architecture, and the two TMs' gravitational accelerations (cf. Eq. (\ref{eq:TMforcesimple}), (\ref{eq:E}), and (\ref{eq:Greal})). For the orbit propagation, it involves not just a single satellite, but a system of nine objects (three satellites and six TMs). This nine-object configuration highlights the unique characteristics of GW detection satellite orbits due to the drag-free control.


\section{Implementation of simulations}\label{sec: implementation}
This section shows how to perform the high-precision simulation of TianQin dynamics (\texttt{TQDYN}). Considering the baseline of $1.7 \times 10^{8}$ m and the displacement induced by the drag-free control of about $10^{-12}$ m, the observable range must be specified with a precision of at least 20 digits, which exceeds the 16-digit precision limit of the double-precision format. For a realistic simulation, we account for the fact that the satellite's attitude is influenced by the directions of incoming light rather than the satellite's instantaneous positions. We implement an approximation to separate the attitude from light-tracing issues, ensuring the simulation remains feasible and that the induced error stays within the acceptable noise limits of the detection band. 

\subsection{High-precision orbits}
The \texttt{TQPOP} (TianQin Quadruple Precision Orbit Propagator) program \cite{zhang2021effect} based on \texttt{MATLAB} has been developed to evaluate the range acceleration between two objects at $<10^{-15}$ m/s$^2$/Hz$^{1/2}$ levels. The quadruple precision data type is applied to all the necessary aspects of the program, including parameter inputs, ephemeris data outputs, reference frame transformations, time conversion, numerical integration, force models, etc. For nearly circular high orbits, the integrator employs the 8th-order embedded Prince-Dormand (DP87) method \cite{Prince1981} with a fixed step size of 50 seconds, corresponding to a Nyquist frequency of $10^{-2}$ Hz. This algorithm achieves a relative truncation error of less than $<10^{-20}$, ensuring more than 20 significant digits of precision in both satellite positions and velocities. Consequently, the estimated range acceleration error is below $<10^{-22}$ m/s$^2$, which is significantly lower than $10^{-15}$ m/s$^2$. The round-off error, resulting from finite numerical precision, is approximately $<10^{-33}$ m/s$^2$/Hz$^{1/2}$, and no longer represents a limiting factor. This eliminates the risk of round-off errors overwhelming gravity field signals, as would occur with double-precision calculations. The \texttt{TQDYN} program \cite{fang2024payload} is designed to compute the nominal control of the TMs and the satellite, while also providing references for the satellite architecture. This application employs the satellite's pure gravity orbits as inputs to calculate attitude dynamics, while disregarding the impact of attitude back to the orbits. 

By integrating \texttt{TQDYN} with \texttt{TQPOP}, we develop an enhanced version of \texttt{TQDYN}, operating in quadruple precision, and capable of generating the dynamics of three satellites, thus addressing the issues associated with DFPC effects on satellite orbits. For simplicity, we no longer differentiate between \texttt{TQDYN} and its enhanced quadruple version, referring to both uniformly as \texttt{TQDYN}. Initially, \texttt{TQDYN} acquires the initial orbit states of three satellites, and then, employing the orbit-attitude relationship described in Sec. \ref{sec:OAdependence1}, calculates the dynamic states, including the attitudes and orbits of two TMs and the orientations of two MOSAs. Subsequently, the program utilizes celestial force models to calculate the gravitational acceleration affecting the 9-body system (comprising 6 TMs and 3 satellites) and integrates it with the drag-free scheme outlined in Sec. \ref{sec:OAdependence2} for orbit propagation. To efficiently facilitate propagation, we creatively transformed the dynamic states of three satellites into a matrix, enabling simultaneous integration through MATLAB's matrix operations at each propagation step. Although the whole approach extends the computation time, a significant advantage is that it allows for the simulation of the entire trio of satellites within the constellation simultaneously.

\subsection{Satellite pointing}\label{sec:decoupledComputation}
The TianQin satellite's orientation is governed by the two incoming laser beams. In this scenario, a more realistic model needs to utilize the local constellation plane for defining the satellite attitudes (cf. Eqs. (\ref{eq:realpointing})). The pointing calculation faces challenges due to the need for iterative solving of the implicit Eqs. (\ref{eq:lightTravelTime}), particularly when the effects of general relativity are neglected in light propagation (cf. \cite{chauvineau2005relativistic, otto2015time}). The iterations at each integration step render the orbit propagation program inefficient and time-consuming. Fortunately, some approximations \cite{han2018point} can be used in the simulation to mitigate these issues.

For instance, setting Sat$_i$ as the local satellite. The positions of the distant satellites $\vec{r}_j(t_0-\tau_{ij'})$ and $\vec{r}_k(t_0-\tau_{ik'})$ can be approximated using Taylor expansions of $\vec{r}_j(t)$, $\vec{r}_k(t)$ at $t_0$, which read as
\begin{equation}\label{eq:Taylorextension1}
\begin{aligned}
    \vec{r}_j(t_0-\tau_{ij'}) = \vec{r}_j(t_0) &- \dot{\vec{r}}_j(t_0)\tau_{ij'} \\ &+ \frac{1}{2}\ddot{\vec{r}}_j(t_0){\tau_{ij'}}^2 + \mathcal{O}({\tau_{ij'}}^3),  
\end{aligned}
\end{equation}
and
\begin{equation}\label{eq:Taylorextension2}
\begin{aligned}
    \vec{r}_k(t_0-\tau_{ik'}) = \vec{r}_k(t_0) &- \dot{\vec{r}}_k(t_0)\tau_{ik'} \\ &+ \frac{1}{2}\ddot{\vec{r}}_k(t_0){\tau_{ik'}}^2 + \mathcal{O}({\tau_{ik'}}^3). 
\end{aligned}
\end{equation}
Using Eqs. (\ref{eq:realpointing}) in conjunction with Eqs. (\ref{eq:Taylorextension1}) and (\ref{eq:Taylorextension2}), while disregarding higher-order terms beyond the third order. The approximations of the $\vec{n}_{ij'}$ and $\vec{n}_{ik'}$ read
\begin{equation}\label{eq:pointing1}
\begin{aligned}
    \vec{n}_{ij'} \;{\approx}\; \frac{\vec{r}_j(t_0) - \vec{r}_i(t_0)}{c\tau_{ij'}} - \frac{\dot{\vec{r}}_j(t_0)}{c} + \frac{\ddot{\vec{r}}_j(t_0)\tau_{ij'}}{2c},\\
    \vec{n}_{ik'} \;{\approx}\; \frac{\vec{r}_k(t_0) - \vec{r}_i(t_0)}{c\tau_{ik'}} - \frac{\dot{\vec{r}}_k(t_0)}{c} + \frac{\ddot{\vec{r}}_k(t_0)\tau_{ik'}}{2c}.
\end{aligned}
\end{equation}

Since the primary aim of this part simulation is to calculate the pointings, not the accurate value of the LTT itself (a more precise determination of the LTTs, for light-path simulation, can be found in Sec. \ref{sec:simulatedLPS}). Hence, a convenient and yet reliable approximation involves simplifying it as
\begin{equation}\label{eq:pointing2}
\begin{aligned}
    \tau_{ij'} \approx \frac{|\vec{{r}_j}(t_0) - \vec{{r}_i}(t_0)|}{c},\\
    \tau_{ik'} \approx \frac{|\vec{{r}_k}(t_0) - \vec{{r}_i}(t_0)|}{c},
\end{aligned}
\end{equation}
which will introduce a time error around 10$^{-7}$ s in TianQin's scenario. To combine Eqs. (\ref{eq:pointing1}) with Eqs. (\ref{eq:pointing2}) and Eqs. (\ref{eq:geometrypointing}), the $\vec{n}_{ij'}$ and $\vec{n}_{ik'}$ can be represented as follow
\begin{equation}\label{eq:realpointingapproximation1}
\begin{aligned}
    \vec{n}_{ij'} \approx \frac{\vec{r}_j(t_0) - \vec{r}_i(t_0)}{|\vec{{r}_j}(t_0)-\vec{{r}_i}(t_0)|} & -  \frac{\dot{\vec{r}}_j(t_0)}{c} \\& + \frac{\ddot{\vec{r}}_j(t_0)|\vec{{r}_j}(t_0) - \vec{{r}_i}(t_0)|}{2c^2},
\end{aligned}
\end{equation}
and
\begin{equation}\label{eq:realpointingapproximation2}
\begin{aligned}
    \vec{n}_{ik'} \approx \frac{\vec{r}_k(t_0) - \vec{r}_i(t_0)}{|\vec{{r}_k}(t_0) - \vec{{r}_i}(t_0)|} & - \frac{\dot{\vec{r}}_k(t_0)}{c} \\& + \frac{\ddot{\vec{r}}_k(t_0)|\vec{{r}_k}(t_0) - \vec{{r}_i}(t_0)|}{2c^2}.
\end{aligned}
\end{equation}

In Eqs. (\ref{eq:realpointingapproximation1}) and (\ref{eq:realpointingapproximation2}), the attitude determined by the local constellation plane can be decoupled from the implicit Eqs. (\ref{eq:lightTravelTime}). To validate the approximation method, we compute the pointing directions using various attitude determination methods, i.e., the iteration method, the approximation method, and the global constellation attitude defined by Eqs. (\ref{eq:geometrypointing}).

\begin{figure}[ht]
  \centering
  \includegraphics[width=0.48\textwidth]{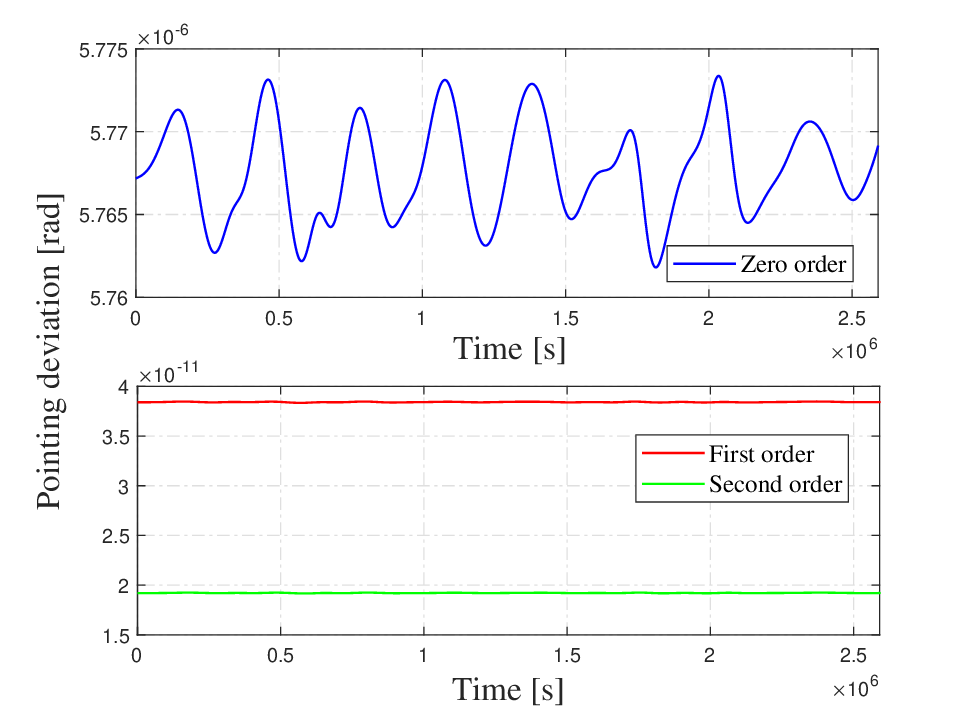}
    \caption{\label{fig:DeviationofPoint} Comparison of different strategies for calculating satellite pointing. In the case of TianQin, the global constellation method will introduce a pointing error of $10^{-6}$ rad (upper plot), while the approximation method for pointing calculation shows accuracy errors on the order of $10^{-11}$ rad (lower plot).}
\end{figure}

The pointing directions calculated through iteration are used as a benchmark for its high accuracy. As illustrated in Fig. \ref{fig:DeviationofPoint}, the global constellation attitude method yields a pointing error on the order of 10$^{-6}$ rad (see the upper plot). On the other hand, the approximation method, which employs Taylor series expansions to the first and second orders, results in a significantly lower pointing error on the order of $10^{-11}$ rad, which is better than the DC 10 nrad pointing error for the GW detection satellite pointing requirements. The lower plot illustrates that an increase in orders does not lead to a significant enhancement in accuracy. Therefore, our program utilizes the first-order approximation for simulation purposes.

Furthermore, another primary objective of this study is to assess the effects of range disturbances induced by the DFPC in the frequency domain, where attitude approximation errors may also translate into these range disturbances. The single-link displacement noise requirement is specified by
\begin{equation} \label{eq:req}
\begin{aligned}
S^{1/2}_\mathrm{req} \le 0.3 \, \frac{\mathrm{pm}}{\sqrt{\mathrm{Hz}}} & \sqrt{1+\left(\frac{7\mathrm{mHz}}{f}\right)^{4}},
\end{aligned}
\end{equation}
where 0.1 mHz $\leq$ $f$ $\leq$ 1 Hz. This is allocated from the total noise requirement  \cite{Luo2016}. 
\begin{figure}[ht]
    \centering
    \includegraphics[width=0.48\textwidth]{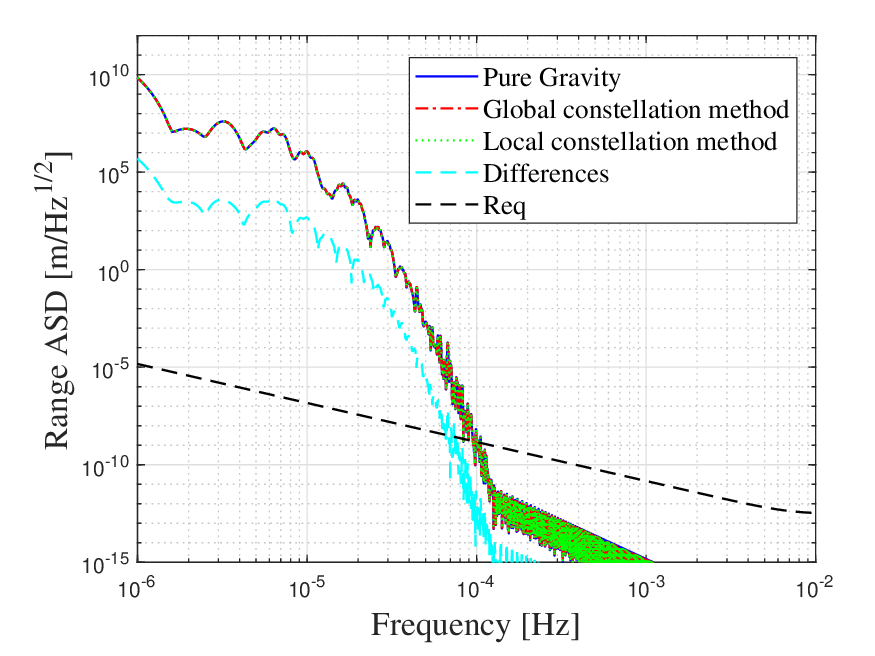}
    \caption{\label{fig:ASDPureGraDFPC} Comparative analysis of the range ASD of different satellite orbits, including the gravity orbits and two drag-free control orbits calculated through global or local constellation approximation methods.}
\end{figure}

We evaluated the range amplitude spectral density (ASD) of single-link TianQin satellites under drag-free control using two distinct attitude determination methods and compared the results with pure gravity satellites. In Fig. \ref{fig:ASDPureGraDFPC}, the solid blue line represents the range ASD of the gravity orbits. The red and green dashed curves represent the range ASD of the drag-free control orbits calculated using global or local constellation approximation methods. Two primary conclusions emerge from the figure. Firstly, for the drag-free satellite, disturbances caused by drag-free will be negligible within the detection frequency range. Secondly, the differences due to different attitude determination techniques fall beneath the noise threshold, as illustrated by the cyan curve. These results reinforce the reliability of calculating the \texttt{TQDYN} through the pointing approximation of the local constellation plane.


\section{Simulation results}\label{sec: selfGraDYN}
This section presents the simulation outcomes of the \texttt{TQDYN}. First, it simulates the detailed coupled dynamics of the constellation for TianQin without self-gravity, and the results show no magnitude difference compared to our previous work under the assumption of orbit-attitude decoupling. Second, the DC self-gravity accelerations on the TMs are simulated to assess the impact on electrostatic control accelerations and constellation stability. The results also help to set restrictions on the self-gravity design. Third, the light path between the two distant TMs can be simulated with a precision of pm/Hz$^{1/2}$ for spectral analysis. It suggests that for split measurement in future TianQin raw data computations, the closed-loop dynamics simulation can be separated from the high-precision orbit propagation.

\subsection{Coupled dynamics in geocentric orbits}\label{sec:coupled dynamics}
\begin{figure}[ht]
  \centering
  \includegraphics[width=0.48\textwidth]{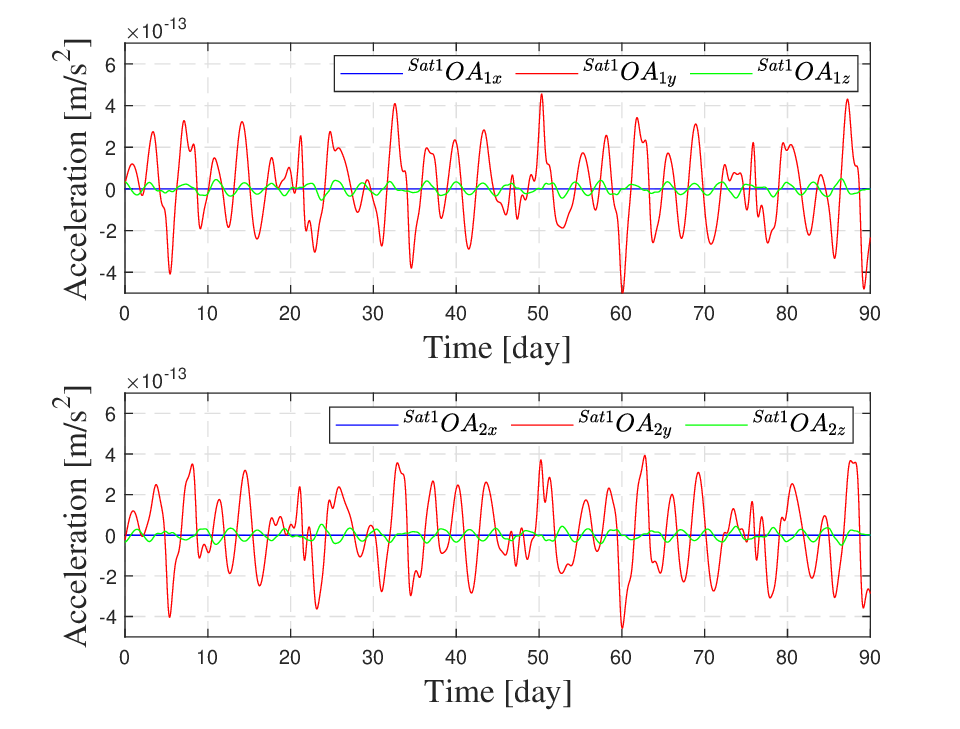}
\caption{\label{fig:ESEarthMoon} Nominal electrostatic control acceleration required for the two TMs with orbit-attitude coupled dynamics. }
\end{figure}

To better demonstrate the effect of the gravitational field on the satellite and TMs in the geocentric orbits with the coupled constellation dynamics, we exclude the influence of self-gravity. The electrostatic suspension controls on the TMs with the geocentric orbit are simulated in the \texttt{TQDYN}. The magnitudes of these electrostatic control accelerations and angular acceleration are around 10$^{-13}$ m/s$^2$ and 10$^{-13}$ rad/s$^2$, which correspond to our previous calculations. Due to the consistency in the magnitude of the results, only the key schematic of the electrostatic control acceleration is shown in Fig. \ref{fig:ESEarthMoon} for simplicity.

In the realm of satellite orbits, because the locations of the two TMs and the satellite's CoM are in close proximity, the differential acceleration due to gravity gradients, along with inertial accelerations, remains below 10$^{-12}$ m/s$^2$. The drag-free acceleration $\vec{G}_i$ experienced by the satellite is approximately 10$^{-13}$ m/s$^2$. Therefore, even with the two sensitive axes drag-free control strategy, the satellite position will remain near its pure gravity orbit, with deviations less than a meter over three months. The results indicate that even with a stronger gravity gradient than that of the heliocentric orbit, the geocentric orbit remains feasible in principle for the two sensitive axes drag-free control strategy. Furthermore, the assumption of using the CoM of the satellite as the inertial reference point proves to be valuable for optimizing mechanical parameters, which provides essential guidance for architecture design. 

\subsection{Limitations on self-gravity}\label{sec:limitation on selfGra}

In this section, a comprehensive analysis of the direction and magnitude of the DC self-gravity effects on the satellite and the TMs through suspension and drag-free control will be presented. It can be constrained in two main aspects: the restriction of 10$^{-10}$ m/s$^2$ for the electrostatic control accelerations that act on the TM and the requirements to maintain the stability of the constellation with the variation of the breathing angle below $\pm$ 0.1$^\circ$ \cite{Ye2019}.

\subsubsection{Electrostatic control acceleration limitation}
The celestial bodies combined with the inertial acceleration on the TMs are estimated to be compensated for by electrostatic accelerations at the level of 10$^{-13}$ m/s$^2$ in figure \ref{fig:ESEarthMoon}. Meanwhile, the self-gravity, which is considered as the main DC accelerations on the TMs, also needs to be compensated for by the electrostatic accelerations. Given that the control acceleration is solely dependent on the differential acceleration between the two TMs (cf. Eqs. (\ref{eq:E1}) and (\ref{eq:E2})), equalizing the orientations and magnitudes of the self-gravity on the TMs is an efficient strategy to reduce actuation. Theoretically, if the self-gravity is identical for both TMs, additional electrostatic control adjustments are unnecessary. However, the actual design must be considered. To provide a more detailed simulation, we incorporate the most recent self-gravity design for TianQin \cite{chen23selfgravity}. Note that these parameters are intended solely for dynamic analysis and do not signify the final design (see Table \ref{tab:SelfGra}). 
\begin{table}[htb]
\caption{\label{tab:SelfGra} The primary self-gravity design in the corresponding TM frames. }
\begin{ruledtabular}
\begin{tabular}{ccc}
Symbols & Values
\\ 
\hline
$\mathrm{X_{TM_1}}$  & $4.94^{-11}$ m/s$^2$ \\
$\mathrm{X_{TM_2}}$  & $5.20^{-11}$ m/s$^2$ \\
\hline
$\mathrm{Y_{TM_1}}$  & $6.22^{-12}$ m/s$^2$ \\
$\mathrm{Y_{TM_2}}$  & $5.65^{-12}$ m/s$^2$ \\
\hline
$\mathrm{Z_{TM_1}}$  & $1.10^{-10}$ m/s$^2$ \\
$\mathrm{Z_{TM_2}}$  & $-5.72^{-11}$ m/s$^2$ \\
\end{tabular}
\end{ruledtabular}
\end{table}

\begin{figure}[ht]
  \centering
  \includegraphics[width=0.48\textwidth]{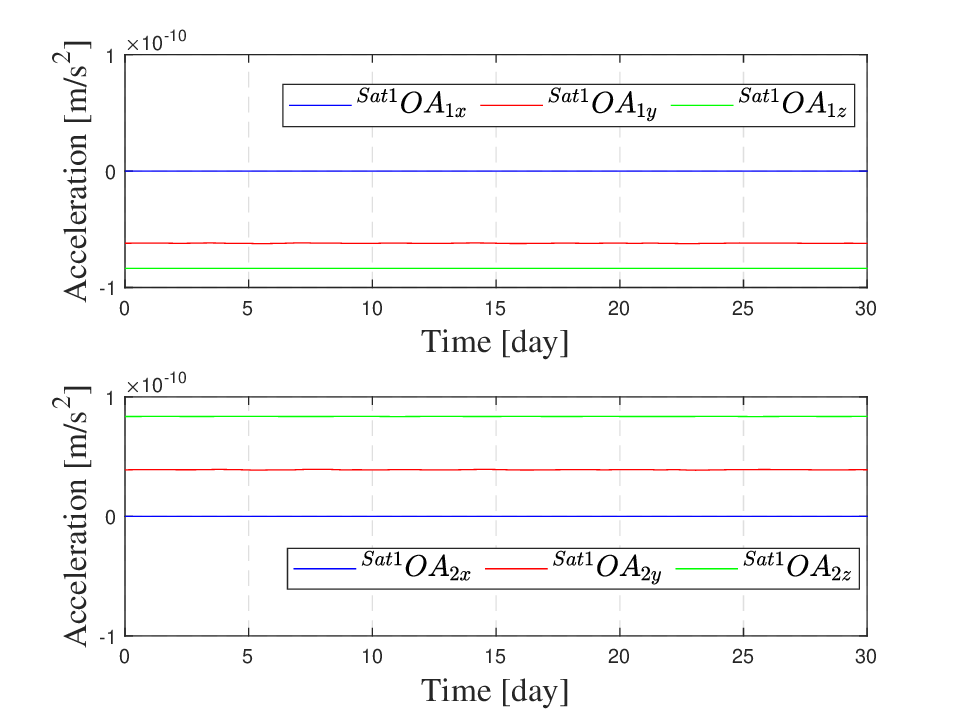}
    \caption{\label{fig:ESselfGravity} Suspension control acceleration on two TMs with the current self-gravity design over one month. }
\end{figure}

The one-month propagation of the satellite dynamics results in the electrostatic control on TM can be seen in Fig. \ref{fig:ESselfGravity}. The gravity and inertial accelerations that need to be compensated by electrostatic control, are minuscule ($10^{-13}$ m/s$^2$) in comparison to the DC self-gravity ($10^{-10}$ m/s$^2$) that also need to be compensated by the control, thereby exerting negligible impact in this figure. To conclude, the current configuration of self-gravity meets the DFPC standards for electrostatic acceleration constraints, with differential acceleration between the two TMs less than $10^{-10}$ m/s$^2$. 

\subsubsection{Constellation stability}
As the controls along the sensitive axes are implemented solely through satellite actuation. The DC self-gravity will exert an influence on the orbits of the satellites, thereby impacting the entirety of the constellation. The main components of the actuation on the satellite are associated with the common part of self-gravity accelerations of the two TMs. Once the differential acceleration is neutralized through electrostatic suspension (refer to Eqs. (\ref{eq:E1}) and (\ref{eq:E2})), the common acceleration (see \ref{eq:Greal}) will be achieved using drag-free control.

To examine how drag-free control affects satellite orbits in the presence of self-gravity, we explore situations where the self-gravity accelerations on both TMs are about $10^{-9}$ m/s$^2$, with these accelerations aligned in the same direction according to the satellite's coordinate system. We examined the scenario along all three axes of the satellite coordinate, considering both positive and negative directions. Under these conditions, self-gravity has the most significant impact on the orbit. It shows that the orbits are most sensitive to self-gravity in the flight direction, which can reach the orders of 100 km deviation from the origin pure gravity orbits over 3 months. However, in the aspect of the constellation, the breathing angles exhibit minimal alteration, maintaining a variance well below $\pm$ 0.1$^\circ$.

\begin{figure}[ht]
  \centering
  \includegraphics[width=0.48\textwidth]{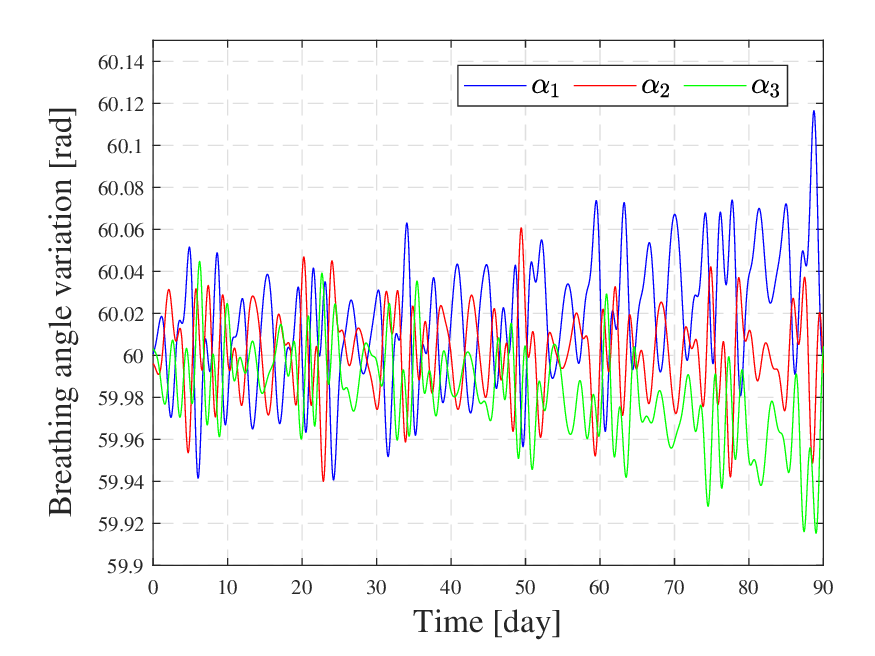}
\caption{\label{fig:deviation} Breathing angle variation over three months, with the extreme case of self-gravity. The breathing angles show tendency to diverge.}
\end{figure}

Furthermore, some relatively extreme cases are simulated, where the common self-gravity directions are aligned with or against the flight direction across the three satellites. To maximize the disruption of the constellation configuration, we set the self-gravity of the different satellites to pull apart in the direction of the inter-satellite distances. For Sat$_1$ and Sat$_2$, the self-gravity directions are along the positive and negative flight directions, respectively. However, for Sat$_3$, the self-gravity direction is along the positive flight direction (for Sat$_3$, the positive and negative directions have no difference in impact on the constellation configuration). In this scenario, the breathing angles of the constellation show a clear tendency to drift apart. The three-month simulation results are shown in Fig. \ref{fig:deviation}. It shows that with the 100 km deviation from the origin gravity orbits for each satellite, the constellation may diverge and exceed the requirement of 60$\pm0.1^\circ$ near the end of the 3 months. 

In summary, having the same self-gravity along the flight directions for all three satellites is favorable for maintaining the constellation stability, despite some deviations from the ideal gravity orbits. Alternatively, the self-gravity of each satellite should be oriented away from the flight direction, with a preferred perpendicular alignment to the constellation plane, to minimize its impact on the orbit. 


\subsection{Spectral analysis}\label{sec:simulatedLPS}
For simulating the raw dynamic data of TMs and satellites in the split measurement, integrating closed-loop simulations (i.e., the instrument noises) with high-precision orbit propagation (i.e., the celestial gravitational noises) is required. However, the orbit computation part in closed-loop control (often simulated by double precision) introduces significant round-off errors. Such a combination between the constellation-level and the instrument-level is relatively difficult to achieve, requiring significant computational resources, being time-consuming, and having low efficiency. 

Fortunately, the evolutions of the nominal states, also known as the working points of the closed-loop control system (e.g., the nominal satellite/TM/MOSA orbits, attitudes, and dynamics, etc.) are mainly associated with the celestial environment and control strategies, which can be simulated by \texttt{TQDYN}. If the effects of nominal attitude dynamics result in ranging noises between two TMs being outside the target frequency band, then the attitude noises will be influenced solely by the jitters from the closed-loop simulation. Thereby permitting independent simulations of these two.

\begin{figure}[ht]
    \centering
    \includegraphics[width=0.48\textwidth]{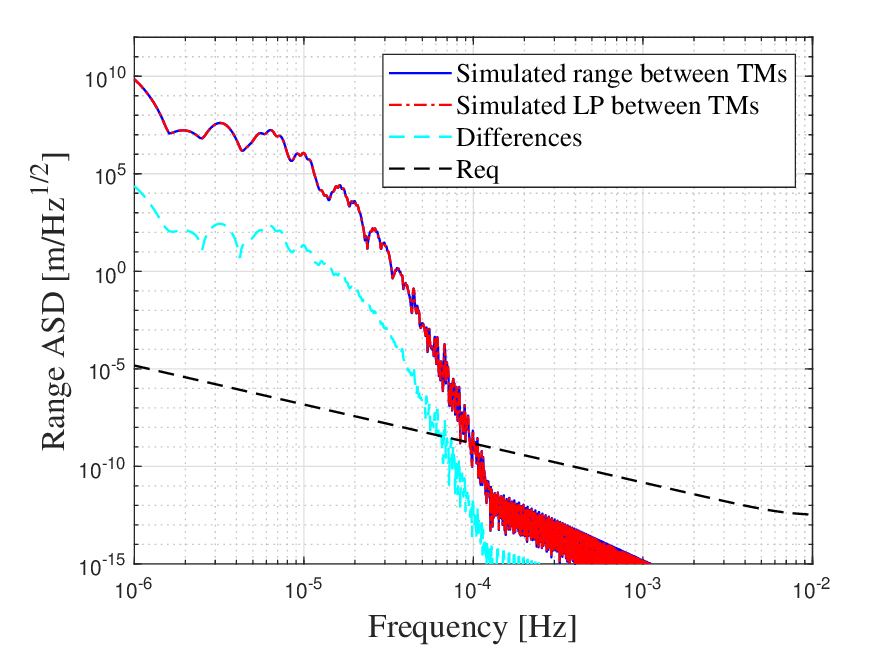}
    \caption{\label{fig:ASDRangeLP} Range ASD of the real light path and the instantaneous range for $Sat_1$'s TM to corresponding $Sat_3$'s TM.}
\end{figure}

First, as both satellites and TMs are managed by the nominal DFPC, our program can simulate the nominal attitude dynamics and control actuations for each satellite and TM. We show the simulation of the light path along a single arm link and analyze it in the range ASD to estimate the range noises. Simulating the light path involves computing the distance between the local TM at time $t_0$ and the remote TM at delayed time ${t_0}-\tau$, where $\tau$ represents the LTT between the two spatially separated TMs. The trajectories of the local TM and its distant counterpart can be modeled using \texttt{TQDYN}. Subsequently, Eqs. (\ref{eq:lightTravelTime}) can be employed with an iterative approach to achieve high accuracy (up to 10$^{-34}$ m) in the light path simulation \cite{wang2024point}. The ASD curves for the instantaneous range of the drag-free TMs and the light-path range show no differences within the detection band, and both precisions exceed the pm/Hz$^{1/2}$ level, as illustrated in Fig. \ref{fig:ASDRangeLP}. The long-range light path can therefore be combined with TQTDI \cite{zheng2023doppler} for further split interferometry simulation.

\begin{figure}[ht]
    \centering
    \includegraphics[width=0.48\textwidth]{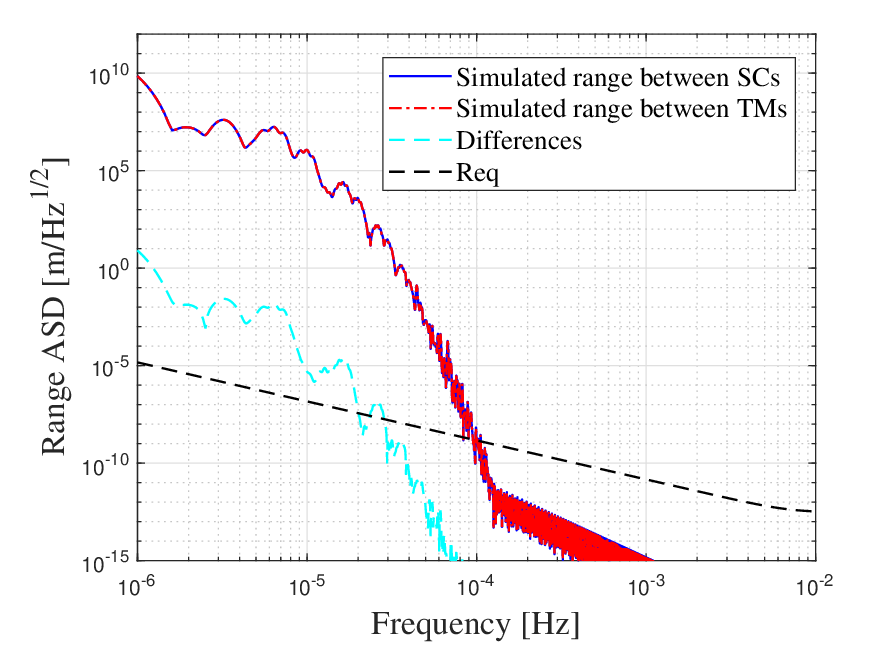}
    \caption{\label{fig:ASDSCandTM} Range ASD of the $Sat_1$ to $Sat_3$'s instantaneous range and the $Sat_1$'s TM to corresponding $Sat_3$'s TM instantaneous range. }
\end{figure}

Next, the simulation of pm/Hz$^{1/2}$ level long-range light path allows us to assess whether the attitude impact falls within the detection frequency band. We simulate both the satellite-to-satellite instantaneous range ASD and the corresponding TM-to-TM instantaneous range ASD, as demonstrated in Fig. \ref{fig:ASDSCandTM}. It shows that the impacts of the nominal drag-free and suspension control disturbances on satellites and TMs are below the requirement within the detection band. Therefore, for TianQin dynamics simulations, it can be assumed that the closed-loop control dynamics simulation is independent from the high-precision orbit propagation. This enables a more focused analysis of each simulation's features with computational efficiency. It is noticeable that the residual noises from the closed-loop dynamics are not considered in this simulation. In future studies, emphasis will be placed on its coupling contribution to the TTL noises.


\section{Conclusion and discussion}\label{sec: conclusion}
In this study, we propose a comprehensive dynamical evolution model that accounts for the mutual coupling relationships between the six TMs and the three satellites (both attitudes and orbits), achieving high-precision simulation with the force models of gravity in geocentric orbits and the satellite self-gravity. Using this dynamics model, we initially revalidated the feasibility of the TianQin mission's two TMs and telescope pointing scheme through the comprehensive orbit-attitude coupled dynamics. 

Second, we assess the impact of self-gravity from two perspectives: constellation stability and electrostatic suspension control. We established some criteria for self-gravity design. To help maintain the constellation stability, it is recommended that the DC common self-gravity along the flight direction for the three satellites should be minimized or at least kept close to one another. For the differential self-gravity component, the magnitudes should be maintained at $\sim 10^{-10}$ m/s$^2$ level to mitigate the electrostatic control force noise on the TMs. 

Third, a numerical simulation of the long-range light path between two TMs with pm/Hz$^{1/2}$ precision is presented. This simulation incorporates Earth-Moon system gravitational effects alongside self-gravity to model coupled constellation dynamics, subsequently identifying the light path linking the two TMs once drag-free control is established. The ASD outcomes are consistent with expectations, demonstrating that under the nominal DFPC, the self-gravity and the majority of Earth-Moon gravity disturbances between two TMs remain outside TianQin's detection frequency band above $1\times 10^{-4}$ Hz. Moreover, as the perturbations from the nominal attitudes are not in the detection frequency band, future research on TianQin dynamics simulations can independently simulate the control residual noises from closed-loop control alongside high-precision orbit propagation.

This paper explores the coupled constellation dynamics in the context of space-based GW detection missions, utilizing the high-precision simulations. Some intriguing future research based on this paper remains. For instance, the framework of the high-precision data of orbits and attitudes has been established, which can be further integrated with the closed-loop control noise model in the split measurement to study the effect of TTL. Additionally, the utilization of high-precision orbit propagation with the self-gravity effect compared with the TianQin's access to Global Navigation Satellite Systems enables on-orbit calibration of the satellite self-gravity.


\begin{acknowledgments}
The authors thank Shutao Yang for help with the program, Enrico Canuto, Bobing Ye, and Dezhi Wang for helpful discussions and comments.  X. Z. is supported by the National Key R\&D Program of China (Grant No. 2022YFC2204600 and 2020YFC2201202), NSFC (Grant No. 12373116). 
\end{acknowledgments}


\appendix

\section{DETAILED ATTITUDE DYNAMICS}\label{sec:AttitudeDYN}
In Sec. \ref{sec:OAdependence1}, the dependence of one satellite's nominal attitude on the constellation is deduced. In this appendix, we introduce the relationships between the orbital velocities and accelerations of the three satellites and their corresponding nominal angular velocities and accelerations. First, we introduce that, for example, the coordinate transformation matrix from the $\mathcal{I}$ frame to the ${\mathcal{S}_i}$ frame can be determined by
\begin{equation}\label{eq:trans}
T_I^{S_i}=
\begin{bmatrix}
 \vec X_{S_i}^T \\
 \vec Y_{S_i}^T \\
 \vec Z_{S_i}^T
\end{bmatrix}, 
\end{equation}
where the direction column vectors of ${\mathcal{S}_i}$ are expressed in the $\mathcal{I}$ frame, and the upper right superscript $T$ means the transpose of one column vector. With the rotation matrix time derivative (i.e. Poisson relation) reading \cite{vidano2020lisa}
\begin{equation}\label{eq:omega}
\begin{aligned}
T_I^{S_i} \dot{T}_{S_i}^I &=\begin{bmatrix}
 \vec X_{S_i}^T \\
 \vec Y_{S_i}^T \\
 \vec Z_{S_i}^T
\end{bmatrix} \begin{bmatrix}
\dot{\vec{X}}_{S_i}\;\dot{\vec{Y}}_{S_i}\;\dot{\vec{Z}}_{S_i} 
\end{bmatrix} \\
&= \vec{\omega}_{S_i}\times = \begin{bmatrix}
 0&-\omega_{{S_i}z}&\omega_{{S_i}y} \\
 \omega_{{S_i}z}&0&-\omega_{{S_i}x} \\
 -\omega_{{S_i}y}&\omega_{{S_i}x}&0.
\end{bmatrix}
\end{aligned}
\end{equation}
where $\vec{\omega}_{S_i}$ is the angular velocity vector of the $\mathcal{S}_i$ frame with respect to the inertial frame expressed in the $\mathcal{S}_i$ frame. 

To calculate the angular velocity and acceleration of the satellite in the body frame, it is necessary to determine the velocity and acceleration of the body frame axes. The velocities and range speed of two satellites can be obtained from the orbits, which read 
\begin{equation}\label{eq:OAinput2}
    \dot{\vec{r}}_{ij}={\dot{\vec{R}}_j}-{\dot{\vec{R}}_i},\quad |\dot{\vec{r}}_{ij}|=\dot{\vec{r}}_{ij}\cdot\vec{n}_{ij},
\end{equation}
and the time derivative of the unit vector $\vec{n}_{ij}$ reads
\begin{equation}
    \dot{\vec{n}}_{ij}=\frac{\dot{\vec{r}}_{ij} - |\dot{\vec{r}}_{ij}|\vec{n}_{ij}}{|\vec{r}_{ij}|}.
\end{equation}
The accelerations part follows the same procedure as the velocities; thus it can be written as
\begin{equation}\label{eq:OAinput3}
    \ddot{\vec{r}}_{ij}={\ddot{\vec{R}}_j}-{\ddot{\vec{R}}_i},\quad |\ddot{\vec{r}}_{ij}|=\ddot{\vec{r}}_{ij}\cdot\vec{n}_{ij} + \dot{\vec{r}}_{ij}\cdot\dot{\vec{n}}_{ij} ,
\end{equation}
and the time derivative of the velocity unit vectors $\dot{\vec{n}}_{ij}$ reads
\begin{equation}
    \ddot{\vec{n}}_{ij}=\frac{\ddot{\vec{r}}_{ij} - |\ddot{\vec{r}}_{ij}|\vec{n}_{ij} - 2|\dot{\vec{r}}_{ij}|\dot{\vec{n}}_{ij}}{|\vec{r}_{ij}|}
\end{equation}

The breathing angle variations of this constellation can be determined by the first- and second-time derivatives of Eq. (\ref{eq:breathingAngle}), which reads as
\begin{equation}
    \dot{\alpha}_i=-\frac{\dot{\vec{n}}_{ij} \cdot \vec{n}_{ik} + \vec{n}_{ij} \cdot \dot{\vec{n}}_{ik}}{\sin\alpha_i},
\end{equation}
and 
\begin{equation}
    \ddot{\alpha}_i=-\frac{\ddot{\vec{n}}_{ij}\cdot\vec{n}_{ik} + \vec{n}_{ij}\cdot\ddot{\vec{n}}_{ik} + 2\dot{\vec{n}}_{ij}\cdot\dot{\vec{n}}_{ik} + \dot{\alpha}_i^2\cos{\alpha_i}}{\sin\alpha_i}.
\end{equation}

When the attitudes of the satellites are locked onto the constellation, the angular velocities and accelerations of each satellite can be determined. This enables calculation of the dynamics of the three satellites in relation to one reference frame that lies within the constellation plane, but is not tied to any single satellite. Instead, it is defined by the constellation itself and maintains a relationship to the inertial reference frame. Then we introduce an auxiliary frame called the $p$-frame with the origin at the constellation's incenter $\vec{r}_{\mathrm{inc}}$, and use the super or subscript $p$ to annotate variables related to or expressed in the frame. The $z$-axis of the $p$-frame is
\begin{equation}
    \vec{z}_p = \vec{Z}_{S_i} = \frac{\vec n_{ij} \times \vec n_{ik}}{\sin\alpha_i},
\end{equation}
and the $x$-axis aligns with the projection of $\vec{Z}_I$ onto the constellation plane,
\begin{equation}
    \vec{x}_p=\frac{\cos{\beta}\vec{z}_p-\vec{Z}_I}{\sin\beta}, 
\end{equation}
with
\begin{equation}
    \cos\beta=\vec{Z}_I\cdot\vec{z}_p,
\end{equation}
and $\vec{y}_p$ given by the right-hand rule. Hence, the angular velocity of the $p$-frame, see Eq. (\ref{eq:omega}), is given by
\begin{equation}\label{eq:omega_p}
\vec{\omega}_p=
\begin{bmatrix}
\vec{z}_p\cdot\dot{\vec{y}}_p\\
\vec{x}_p\cdot\dot{\vec{z}}_p\\
\vec{y}_p\cdot\dot{\vec{x}}_p
\end{bmatrix}.
\end{equation}
With the time derivative of the Eq. (\ref{eq:omega_p}), the angular acceleration of the $p$ frame expressed in the $p$ frame can be written as 
\begin{equation}
    \dot{\vec{\omega}}_p=
    \begin{bmatrix}
    \dot{\vec{z}}_p\cdot\dot{\vec{y}}_p+\vec{z}_p\cdot\ddot{\vec{y}}_p\\
    \dot{\vec{x}}_p\cdot\dot{\vec{z}}_p+\vec{x}_p\cdot\ddot{\vec{z}}_p\\
    \dot{\vec{y}}_p\cdot\dot{\vec{x}}_p+\vec{y}_p\cdot\ddot{\vec{x}}_p
    \end{bmatrix}.
\end{equation}

For the satellite attitude dynamics with the $p$ frame. The position of the satellite in the $p$ frame can be read as
\begin{equation}\label{eq:T_I^p}
    ^p\!\vec{R}_i=T_I^p({\vec{R}_i}-{\vec{r}_{\mathrm{inc}}}),
\end{equation}
where $T_I^p$ follows the process of Eq. (\ref{eq:trans}) with $\vec{x}_p$, $\vec{y}_p$, and $\vec{z}_p$ instead. Thus, the $\vec X_{S_i}$ vector of the satellite $i$ is written in the $p$-frame as
\begin{equation}
    {^p\vec{x}_i}=-\frac{^p\vec{R}_i}{|^p\vec{R}_i|},
\end{equation}
and the three satellite $z$ axes are the same as the $\vec{z}_p$, which represents in the $p$-frame as 
\begin{equation}
    ^p\vec{z}_i = \;^p\vec{z}_p = 
    \begin{bmatrix}
        0\\
        0\\
        1
    \end{bmatrix}.
\end{equation}
The $^p\vec{y}_i$ of each satellite follows the right-hand rule.

By definition, the x-y planes of the three satellites all lie within the $p$-plane. As a result, the angular velocity of the satellite $i$ with respect to the $p$ frame only has a nonzero $z$-component, i.e.,
\begin{equation}
    ^p\vec{\omega}_{i/p} = {^p\vec{x}_i} \times {^p\dot{\vec{x}}_i} = \frac{^p\!\dot{\vec{R}}_i\times{^p\!\vec{x}_i}}{|^p\vec{R}_i|}.
\end{equation}

The angular acceleration of the satellite $i$ with respect to the frame $p$ is given by the time derivative of $^p\vec{\omega}_{i/p}$, which reads
\begin{equation}
    ^p\dot{\vec{\omega}}_{i/p} = \frac{2({^p\vec{x}_i}\cdot{^p\!\dot{\vec{R}}_i}){^p\vec{\omega}_{i/p}}-{^p\vec{x}_i}\times{^p\!\ddot{\vec{R}}_i}}{|{^p\!\vec{R}_i}|}.
\end{equation}

Finally, the angular velocity and acceleration of the satellite ${^{S_i}\vec{\omega}_{i}}$, ${^{S_i}\dot{\vec{\omega}}_{i}}$ can be represented as 
\begin{equation}\label{eq:T_p^{S_i}}
    {^{S_i}\vec{\omega}_{i}} = T_p^{S_i}({\vec{\omega}_{p}}+{^p\vec{\omega}_{i/p}}),
\end{equation}
and
\begin{equation}
    {^{S_i}\dot{\vec{\omega}}_{i}} = T_p^{S_i}({\dot{\vec{\omega}}_{p}}+{^p\dot{\vec{\omega}}_{i/p}})+\dot{T}_p^{S_i}({\vec{\omega}_{p}}+{^p\vec{\omega}_{i/p}}),
\end{equation}
where $T_p^{S_i}$ follows the process of Eq. (\ref{eq:trans}) with $^p\vec{x}_i$, $^p\vec{y}_i$, and $^p\vec{z}_i$ instead.

Likewise, similar relations can be derived for the MOSA's target frame $\mathcal{OA}_l$ by rotating ${\mathcal{S}_i}$ about the $\vec Z_{\mathrm{S}_i}$-axis to get the pointing of MOSAs. For the dynamics part, since the MOSAs are aiming to compensate for the breathing angles, the MOSA's rotating velocity and acceleration in the nominal control should be equal to half of the breathing angle variations. For the MOSA rotation velocities and accelerations in the ${\mathcal{S}_i}$ frame, the ${^{S}{\vec{\omega}}_{\mathrm{OA}_{1/2}}}$ read
\begin{equation}
    {^{S}{\vec{\omega}}_{\mathrm{OA}_{1/2}}}=
    \begin{bmatrix}
    0\\
    0\\
    \pm\frac{1}{2}\dot{\alpha}_i
    \end{bmatrix},
\end{equation}
and the ${^{S}\dot{\vec{\omega}}_{\mathrm{OA}_{1/2}}}$ read
\begin{equation}
    {^{S}\dot{\vec{\omega}}_{\mathrm{OA}_{1/2}}}=
    \begin{bmatrix}
    0\\
    0\\
    \pm\frac{1}{2}\ddot{\alpha}_i
    \end{bmatrix}.
\end{equation}

To sum up, all the information for the calculated attitude dynamics can be deduced originally from Eqs. (\ref{eq:OAinput1}), (\ref{eq:OAinput2}) and (\ref{eq:OAinput3}).

\nocite{*}

\bibliography{bibliography}

\providecommand{\noopsort}[1]{}\providecommand{\singleletter}[1]{#1}%
\begin{thebibliography}{33}%
\makeatletter
\providecommand \@ifxundefined [1]{%
 \@ifx{#1\undefined}
}%
\providecommand \@ifnum [1]{%
 \ifnum #1\expandafter \@firstoftwo
 \else \expandafter \@secondoftwo
 \fi
}%
\providecommand \@ifx [1]{%
 \ifx #1\expandafter \@firstoftwo
 \else \expandafter \@secondoftwo
 \fi
}%
\providecommand \natexlab [1]{#1}%
\providecommand \enquote  [1]{``#1''}%
\providecommand \bibnamefont  [1]{#1}%
\providecommand \bibfnamefont [1]{#1}%
\providecommand \citenamefont [1]{#1}%
\providecommand \href@noop [0]{\@secondoftwo}%
\providecommand \href [0]{\begingroup \@sanitize@url \@href}%
\providecommand \@href[1]{\@@startlink{#1}\@@href}%
\providecommand \@@href[1]{\endgroup#1\@@endlink}%
\providecommand \@sanitize@url [0]{\catcode `\\12\catcode `\$12\catcode
  `\&12\catcode `\#12\catcode `\^12\catcode `\_12\catcode `\%12\relax}%
\providecommand \@@startlink[1]{}%
\providecommand \@@endlink[0]{}%
\providecommand \url  [0]{\begingroup\@sanitize@url \@url }%
\providecommand \@url [1]{\endgroup\@href {#1}{\urlprefix }}%
\providecommand \urlprefix  [0]{URL }%
\providecommand \Eprint [0]{\href }%
\providecommand \doibase [0]{https://doi.org/}%
\providecommand \selectlanguage [0]{\@gobble}%
\providecommand \bibinfo  [0]{\@secondoftwo}%
\providecommand \bibfield  [0]{\@secondoftwo}%
\providecommand \translation [1]{[#1]}%
\providecommand \BibitemOpen [0]{}%
\providecommand \bibitemStop [0]{}%
\providecommand \bibitemNoStop [0]{.\EOS\space}%
\providecommand \EOS [0]{\spacefactor3000\relax}%
\providecommand \BibitemShut  [1]{\csname bibitem#1\endcsname}%
\let\auto@bib@innerbib\@empty
\bibitem [{\citenamefont {{J. Luo et al.}}(2016)}]{Luo2016}%
  \BibitemOpen
  \bibfield  {author} {\bibinfo {author} {\bibnamefont {{J. Luo et al.}}},\
  }\bibfield  {title} {\bibinfo {title} {{TianQin}: a space-borne gravitational
  wave detector},\ }\href@noop {} {\bibfield  {journal} {\bibinfo  {journal}
  {Class. Quantum Grav.}\ }\textbf {\bibinfo {volume} {33}},\ \bibinfo {pages}
  {035010} (\bibinfo {year} {2016})}\BibitemShut {NoStop}%
\bibitem [{\citenamefont {Tan}\ \emph {et~al.}(2020)\citenamefont {Tan},
  \citenamefont {Ye},\ and\ \citenamefont {Zhang}}]{Tan2020}%
  \BibitemOpen
  \bibfield  {author} {\bibinfo {author} {\bibfnamefont {Z.}~\bibnamefont
  {Tan}}, \bibinfo {author} {\bibfnamefont {B.}~\bibnamefont {Ye}},\ and\
  \bibinfo {author} {\bibfnamefont {X.}~\bibnamefont {Zhang}},\ }\bibfield
  {title} {\bibinfo {title} {Impact of orbital orientations and radii on
  {TianQin} constellation stability},\ }\href@noop {} {\bibfield  {journal}
  {\bibinfo  {journal} {Int. J. Mod. Phys. D}\ }\textbf {\bibinfo {volume}
  {29}},\ \bibinfo {pages} {2050056} (\bibinfo {year} {2020})}\BibitemShut
  {NoStop}%
\bibitem [{LIS(2017)}]{LISA2017}%
  \BibitemOpen
  \href@noop {} {\bibinfo {title} {{LISA Laser Interferometer Space Antenna, A
  proposal in response to the ESA call for L3 mission concepts}}},\ \bibinfo
  {howpublished} {arXiv:1702.00786} (\bibinfo {year} {2017})\BibitemShut
  {NoStop}%
\bibitem [{\citenamefont {Armano}\ \emph {et~al.}(2019)\citenamefont {Armano},
  \citenamefont {Audley}, \citenamefont {Baird}, \citenamefont {Binetruy},
  \citenamefont {Born}, \citenamefont {Bortoluzzi}, \citenamefont {Castelli},
  \citenamefont {Cavalleri}, \citenamefont {Cesarini}, \citenamefont {Cruise}
  \emph {et~al.}}]{armano2019lisaDFAC}%
  \BibitemOpen
  \bibfield  {author} {\bibinfo {author} {\bibfnamefont {M.}~\bibnamefont
  {Armano}}, \bibinfo {author} {\bibfnamefont {H.}~\bibnamefont {Audley}},
  \bibinfo {author} {\bibfnamefont {J.}~\bibnamefont {Baird}}, \bibinfo
  {author} {\bibfnamefont {P.}~\bibnamefont {Binetruy}}, \bibinfo {author}
  {\bibfnamefont {M.}~\bibnamefont {Born}}, \bibinfo {author} {\bibfnamefont
  {D.}~\bibnamefont {Bortoluzzi}}, \bibinfo {author} {\bibfnamefont
  {E.}~\bibnamefont {Castelli}}, \bibinfo {author} {\bibfnamefont
  {A.}~\bibnamefont {Cavalleri}}, \bibinfo {author} {\bibfnamefont
  {A.}~\bibnamefont {Cesarini}}, \bibinfo {author} {\bibfnamefont
  {A.}~\bibnamefont {Cruise}}, \emph {et~al.},\ }\bibfield  {title} {\bibinfo
  {title} {Lisa pathfinder platform stability and drag-free performance},\
  }\href@noop {} {\bibfield  {journal} {\bibinfo  {journal} {Physical Review
  D}\ }\textbf {\bibinfo {volume} {99}},\ \bibinfo {pages} {082001} (\bibinfo
  {year} {2019})}\BibitemShut {NoStop}%
\bibitem [{\citenamefont {Colpi}\ \emph {et~al.}(2024)\citenamefont {Colpi},
  \citenamefont {Danzmann}, \citenamefont {Hewitson}, \citenamefont
  {Holley-Bockelmann}, \citenamefont {Jetzer}, \citenamefont {Nelemans},
  \citenamefont {Petiteau}, \citenamefont {Shoemaker}, \citenamefont
  {Sopuerta}, \citenamefont {Stebbins} \emph {et~al.}}]{LISA2024}%
  \BibitemOpen
  \bibfield  {author} {\bibinfo {author} {\bibfnamefont {M.}~\bibnamefont
  {Colpi}}, \bibinfo {author} {\bibfnamefont {K.}~\bibnamefont {Danzmann}},
  \bibinfo {author} {\bibfnamefont {M.}~\bibnamefont {Hewitson}}, \bibinfo
  {author} {\bibfnamefont {K.}~\bibnamefont {Holley-Bockelmann}}, \bibinfo
  {author} {\bibfnamefont {P.}~\bibnamefont {Jetzer}}, \bibinfo {author}
  {\bibfnamefont {G.}~\bibnamefont {Nelemans}}, \bibinfo {author}
  {\bibfnamefont {A.}~\bibnamefont {Petiteau}}, \bibinfo {author}
  {\bibfnamefont {D.}~\bibnamefont {Shoemaker}}, \bibinfo {author}
  {\bibfnamefont {C.}~\bibnamefont {Sopuerta}}, \bibinfo {author}
  {\bibfnamefont {R.}~\bibnamefont {Stebbins}}, \emph {et~al.},\ }\bibfield
  {title} {\bibinfo {title} {Lisa definition study report},\ }\href@noop {}
  {\bibfield  {journal} {\bibinfo  {journal} {arXiv preprint arXiv:2402.07571}\
  } (\bibinfo {year} {2024})}\BibitemShut {NoStop}%
\bibitem [{\citenamefont {Chwalla}\ \emph {et~al.}(2020)\citenamefont
  {Chwalla}, \citenamefont {Danzmann}, \citenamefont {{\'A}lvarez},
  \citenamefont {Delgado}, \citenamefont {Fern{\'a}ndez~Barranco},
  \citenamefont {Fitzsimons}, \citenamefont {Gerberding}, \citenamefont
  {Heinzel}, \citenamefont {Killow}, \citenamefont {Lieser} \emph
  {et~al.}}]{chwalla2020optical}%
  \BibitemOpen
  \bibfield  {author} {\bibinfo {author} {\bibfnamefont {M.}~\bibnamefont
  {Chwalla}}, \bibinfo {author} {\bibfnamefont {K.}~\bibnamefont {Danzmann}},
  \bibinfo {author} {\bibfnamefont {M.~D.}\ \bibnamefont {{\'A}lvarez}},
  \bibinfo {author} {\bibfnamefont {J.~E.}\ \bibnamefont {Delgado}}, \bibinfo
  {author} {\bibfnamefont {G.}~\bibnamefont {Fern{\'a}ndez~Barranco}}, \bibinfo
  {author} {\bibfnamefont {E.}~\bibnamefont {Fitzsimons}}, \bibinfo {author}
  {\bibfnamefont {O.}~\bibnamefont {Gerberding}}, \bibinfo {author}
  {\bibfnamefont {G.}~\bibnamefont {Heinzel}}, \bibinfo {author} {\bibfnamefont
  {C.}~\bibnamefont {Killow}}, \bibinfo {author} {\bibfnamefont
  {M.}~\bibnamefont {Lieser}}, \emph {et~al.},\ }\bibfield  {title} {\bibinfo
  {title} {Optical suppression of tilt-to-length coupling in the lisa long-arm
  interferometer},\ }\href@noop {} {\bibfield  {journal} {\bibinfo  {journal}
  {Physical Review Applied}\ }\textbf {\bibinfo {volume} {14}},\ \bibinfo
  {pages} {014030} (\bibinfo {year} {2020})}\BibitemShut {NoStop}%
\bibitem [{\citenamefont {Paczkowski}\ \emph {et~al.}(2022)\citenamefont
  {Paczkowski}, \citenamefont {Giusteri}, \citenamefont {Hewitson},
  \citenamefont {Karnesis}, \citenamefont {Fitzsimons}, \citenamefont
  {Wanner},\ and\ \citenamefont {Heinzel}}]{paczkowski2022postprocessing}%
  \BibitemOpen
  \bibfield  {author} {\bibinfo {author} {\bibfnamefont {S.}~\bibnamefont
  {Paczkowski}}, \bibinfo {author} {\bibfnamefont {R.}~\bibnamefont
  {Giusteri}}, \bibinfo {author} {\bibfnamefont {M.}~\bibnamefont {Hewitson}},
  \bibinfo {author} {\bibfnamefont {N.}~\bibnamefont {Karnesis}}, \bibinfo
  {author} {\bibfnamefont {E.}~\bibnamefont {Fitzsimons}}, \bibinfo {author}
  {\bibfnamefont {G.}~\bibnamefont {Wanner}},\ and\ \bibinfo {author}
  {\bibfnamefont {G.}~\bibnamefont {Heinzel}},\ }\bibfield  {title} {\bibinfo
  {title} {Postprocessing subtraction of tilt-to-length noise in lisa},\
  }\href@noop {} {\bibfield  {journal} {\bibinfo  {journal} {Physical Review
  D}\ }\textbf {\bibinfo {volume} {106}},\ \bibinfo {pages} {042005} (\bibinfo
  {year} {2022})}\BibitemShut {NoStop}%
\bibitem [{\citenamefont {Heisenberg}\ \emph {et~al.}(2023)\citenamefont
  {Heisenberg}, \citenamefont {Inchausp{\'e}}, \citenamefont {Nam},
  \citenamefont {Sauter}, \citenamefont {Waibel},\ and\ \citenamefont
  {Wass}}]{heisenberg2023lisa}%
  \BibitemOpen
  \bibfield  {author} {\bibinfo {author} {\bibfnamefont {L.}~\bibnamefont
  {Heisenberg}}, \bibinfo {author} {\bibfnamefont {H.}~\bibnamefont
  {Inchausp{\'e}}}, \bibinfo {author} {\bibfnamefont {D.~Q.}\ \bibnamefont
  {Nam}}, \bibinfo {author} {\bibfnamefont {O.}~\bibnamefont {Sauter}},
  \bibinfo {author} {\bibfnamefont {R.}~\bibnamefont {Waibel}},\ and\ \bibinfo
  {author} {\bibfnamefont {P.}~\bibnamefont {Wass}},\ }\bibfield  {title}
  {\bibinfo {title} {Lisa dynamics and control: Closed-loop simulation and
  numerical demonstration of time delay interferometry},\ }\href@noop {}
  {\bibfield  {journal} {\bibinfo  {journal} {Physical Review D}\ }\textbf
  {\bibinfo {volume} {108}},\ \bibinfo {pages} {122007} (\bibinfo {year}
  {2023})}\BibitemShut {NoStop}%
\bibitem [{\citenamefont {Wanner}\ \emph {et~al.}(2024)\citenamefont {Wanner},
  \citenamefont {Shah}, \citenamefont {Staab}, \citenamefont {Wegener},\ and\
  \citenamefont {Paczkowski}}]{wanner2024depth}%
  \BibitemOpen
  \bibfield  {author} {\bibinfo {author} {\bibfnamefont {G.}~\bibnamefont
  {Wanner}}, \bibinfo {author} {\bibfnamefont {S.}~\bibnamefont {Shah}},
  \bibinfo {author} {\bibfnamefont {M.}~\bibnamefont {Staab}}, \bibinfo
  {author} {\bibfnamefont {H.}~\bibnamefont {Wegener}},\ and\ \bibinfo {author}
  {\bibfnamefont {S.}~\bibnamefont {Paczkowski}},\ }\bibfield  {title}
  {\bibinfo {title} {In-depth modeling of tilt-to-length coupling in lisa’s
  interferometers and tdi michelson observables},\ }\href@noop {} {\bibfield
  {journal} {\bibinfo  {journal} {Physical Review D}\ }\textbf {\bibinfo
  {volume} {110}},\ \bibinfo {pages} {022003} (\bibinfo {year}
  {2024})}\BibitemShut {NoStop}%
\bibitem [{\citenamefont {Dam}(2022)}]{dam2022simulations}%
  \BibitemOpen
  \bibfield  {author} {\bibinfo {author} {\bibfnamefont {Q.~N.}\ \bibnamefont
  {Dam}},\ }\emph {\bibinfo {title} {Simulations and associated data analysis
  for realistic LISA configuration}},\ \href@noop {} {Ph.D. thesis},\ \bibinfo
  {school} {Universite Paris Cit{\'e}} (\bibinfo {year} {2022})\BibitemShut
  {NoStop}%
\bibitem [{\citenamefont {Inchausp{\'e}}\ \emph {et~al.}(2022)\citenamefont
  {Inchausp{\'e}}, \citenamefont {Hewitson}, \citenamefont {Sauter},\ and\
  \citenamefont {Wass}}]{inchauspe2022new}%
  \BibitemOpen
  \bibfield  {author} {\bibinfo {author} {\bibfnamefont {H.}~\bibnamefont
  {Inchausp{\'e}}}, \bibinfo {author} {\bibfnamefont {M.}~\bibnamefont
  {Hewitson}}, \bibinfo {author} {\bibfnamefont {O.}~\bibnamefont {Sauter}},\
  and\ \bibinfo {author} {\bibfnamefont {P.}~\bibnamefont {Wass}},\ }\bibfield
  {title} {\bibinfo {title} {{New LISA dynamics feedback control scheme:
  Common-mode isolation of test mass control and probes of test-mass
  acceleration}},\ }\href@noop {} {\bibfield  {journal} {\bibinfo  {journal}
  {Physical Review D}\ }\textbf {\bibinfo {volume} {106}},\ \bibinfo {pages}
  {022006} (\bibinfo {year} {2022})}\BibitemShut {NoStop}%
\bibitem [{\citenamefont {Fang}\ \emph {et~al.}(2024)\citenamefont {Fang},
  \citenamefont {Zhang}, \citenamefont {Fu},\ and\ \citenamefont
  {Li}}]{fang2024payload}%
  \BibitemOpen
  \bibfield  {author} {\bibinfo {author} {\bibfnamefont {Y.}~\bibnamefont
  {Fang}}, \bibinfo {author} {\bibfnamefont {X.}~\bibnamefont {Zhang}},
  \bibinfo {author} {\bibfnamefont {F.}~\bibnamefont {Fu}},\ and\ \bibinfo
  {author} {\bibfnamefont {H.}~\bibnamefont {Li}},\ }\bibfield  {title}
  {\bibinfo {title} {Payload architecture and pointing control strategies for
  tianqin},\ }\href@noop {} {\bibfield  {journal} {\bibinfo  {journal}
  {Physical Review D}\ }\textbf {\bibinfo {volume} {109}},\ \bibinfo {pages}
  {062001} (\bibinfo {year} {2024})}\BibitemShut {NoStop}%
\bibitem [{\citenamefont {Armano}\ \emph {et~al.}(2016)\citenamefont {Armano},
  \citenamefont {Audley}, \citenamefont {Auger}, \citenamefont {Baird},
  \citenamefont {Binetruy}, \citenamefont {Born}, \citenamefont {Bortoluzzi},
  \citenamefont {Brandt}, \citenamefont {Bursi}, \citenamefont {Caleno} \emph
  {et~al.}}]{armano2016constraints}%
  \BibitemOpen
  \bibfield  {author} {\bibinfo {author} {\bibfnamefont {M.}~\bibnamefont
  {Armano}}, \bibinfo {author} {\bibfnamefont {H.}~\bibnamefont {Audley}},
  \bibinfo {author} {\bibfnamefont {G.}~\bibnamefont {Auger}}, \bibinfo
  {author} {\bibfnamefont {J.}~\bibnamefont {Baird}}, \bibinfo {author}
  {\bibfnamefont {P.}~\bibnamefont {Binetruy}}, \bibinfo {author}
  {\bibfnamefont {M.}~\bibnamefont {Born}}, \bibinfo {author} {\bibfnamefont
  {D.}~\bibnamefont {Bortoluzzi}}, \bibinfo {author} {\bibfnamefont
  {N.}~\bibnamefont {Brandt}}, \bibinfo {author} {\bibfnamefont
  {A.}~\bibnamefont {Bursi}}, \bibinfo {author} {\bibfnamefont
  {M.}~\bibnamefont {Caleno}}, \emph {et~al.},\ }\bibfield  {title} {\bibinfo
  {title} {Constraints on lisa pathfinder’s self-gravity: design
  requirements, estimates and testing procedures},\ }\href@noop {} {\bibfield
  {journal} {\bibinfo  {journal} {Classical and quantum gravity}\ }\textbf
  {\bibinfo {volume} {33}},\ \bibinfo {pages} {235015} (\bibinfo {year}
  {2016})}\BibitemShut {NoStop}%
\bibitem [{\citenamefont {Martens}\ and\ \citenamefont
  {Joffre}(2021)}]{martens2021trajectory}%
  \BibitemOpen
  \bibfield  {author} {\bibinfo {author} {\bibfnamefont {W.}~\bibnamefont
  {Martens}}\ and\ \bibinfo {author} {\bibfnamefont {E.}~\bibnamefont
  {Joffre}},\ }\bibfield  {title} {\bibinfo {title} {Trajectory design for the
  esa lisa mission},\ }\href@noop {} {\bibfield  {journal} {\bibinfo  {journal}
  {The Journal of the Astronautical Sciences}\ }\textbf {\bibinfo {volume}
  {68}},\ \bibinfo {pages} {402} (\bibinfo {year} {2021})}\BibitemShut
  {NoStop}%
\bibitem [{\citenamefont {Joffre}\ \emph {et~al.}(2021)\citenamefont {Joffre},
  \citenamefont {Wealthy}, \citenamefont {Fernandez}, \citenamefont {Trenkel},
  \citenamefont {Voigt}, \citenamefont {Ziegler},\ and\ \citenamefont
  {Martens}}]{joffre2021lisa}%
  \BibitemOpen
  \bibfield  {author} {\bibinfo {author} {\bibfnamefont {E.}~\bibnamefont
  {Joffre}}, \bibinfo {author} {\bibfnamefont {D.}~\bibnamefont {Wealthy}},
  \bibinfo {author} {\bibfnamefont {I.}~\bibnamefont {Fernandez}}, \bibinfo
  {author} {\bibfnamefont {C.}~\bibnamefont {Trenkel}}, \bibinfo {author}
  {\bibfnamefont {P.}~\bibnamefont {Voigt}}, \bibinfo {author} {\bibfnamefont
  {T.}~\bibnamefont {Ziegler}},\ and\ \bibinfo {author} {\bibfnamefont
  {W.}~\bibnamefont {Martens}},\ }\bibfield  {title} {\bibinfo {title} {Lisa:
  Heliocentric formation design for the laser interferometer space antenna
  mission},\ }\href@noop {} {\bibfield  {journal} {\bibinfo  {journal}
  {Advances in Space Research}\ }\textbf {\bibinfo {volume} {67}},\ \bibinfo
  {pages} {3868} (\bibinfo {year} {2021})}\BibitemShut {NoStop}%
\bibitem [{\citenamefont {Ye}\ \emph {et~al.}(2021)\citenamefont {Ye},
  \citenamefont {Zhang}, \citenamefont {Ding},\ and\ \citenamefont
  {Meng}}]{ye2021eclipse}%
  \BibitemOpen
  \bibfield  {author} {\bibinfo {author} {\bibfnamefont {B.}~\bibnamefont
  {Ye}}, \bibinfo {author} {\bibfnamefont {X.}~\bibnamefont {Zhang}}, \bibinfo
  {author} {\bibfnamefont {Y.}~\bibnamefont {Ding}},\ and\ \bibinfo {author}
  {\bibfnamefont {Y.}~\bibnamefont {Meng}},\ }\bibfield  {title} {\bibinfo
  {title} {Eclipse avoidance in tianqin orbit selection},\ }\href@noop {}
  {\bibfield  {journal} {\bibinfo  {journal} {Physical Review D}\ }\textbf
  {\bibinfo {volume} {103}},\ \bibinfo {pages} {042007} (\bibinfo {year}
  {2021})}\BibitemShut {NoStop}%
\bibitem [{\citenamefont {Zhang}\ \emph {et~al.}(2021)\citenamefont {Zhang},
  \citenamefont {Luo}, \citenamefont {Jiao}, \citenamefont {Ye}, \citenamefont
  {Yuan}, \citenamefont {Cai}, \citenamefont {Gu}, \citenamefont {Mei},\ and\
  \citenamefont {Luo}}]{zhang2021effect}%
  \BibitemOpen
  \bibfield  {author} {\bibinfo {author} {\bibfnamefont {X.}~\bibnamefont
  {Zhang}}, \bibinfo {author} {\bibfnamefont {C.}~\bibnamefont {Luo}}, \bibinfo
  {author} {\bibfnamefont {L.}~\bibnamefont {Jiao}}, \bibinfo {author}
  {\bibfnamefont {B.}~\bibnamefont {Ye}}, \bibinfo {author} {\bibfnamefont
  {H.}~\bibnamefont {Yuan}}, \bibinfo {author} {\bibfnamefont {L.}~\bibnamefont
  {Cai}}, \bibinfo {author} {\bibfnamefont {D.}~\bibnamefont {Gu}}, \bibinfo
  {author} {\bibfnamefont {J.}~\bibnamefont {Mei}},\ and\ \bibinfo {author}
  {\bibfnamefont {J.}~\bibnamefont {Luo}},\ }\bibfield  {title} {\bibinfo
  {title} {Effect of earth-moon’s gravity on tianqin’s range acceleration
  noise},\ }\href@noop {} {\bibfield  {journal} {\bibinfo  {journal} {Physical
  Review D}\ }\textbf {\bibinfo {volume} {103}},\ \bibinfo {pages} {062001}
  (\bibinfo {year} {2021})}\BibitemShut {NoStop}%
\bibitem [{\citenamefont {Luo}\ and\ \citenamefont
  {Zhang}(2022)}]{luo2022effect}%
  \BibitemOpen
  \bibfield  {author} {\bibinfo {author} {\bibfnamefont {C.}~\bibnamefont
  {Luo}}\ and\ \bibinfo {author} {\bibfnamefont {X.}~\bibnamefont {Zhang}},\
  }\bibfield  {title} {\bibinfo {title} {Effect of earth-moon’s gravity on
  tianqin’s range acceleration noise. ii. impact of orbit selection},\
  }\href@noop {} {\bibfield  {journal} {\bibinfo  {journal} {Physical Review
  D}\ }\textbf {\bibinfo {volume} {105}},\ \bibinfo {pages} {102007} (\bibinfo
  {year} {2022})}\BibitemShut {NoStop}%
\bibitem [{\citenamefont {Jiao}\ and\ \citenamefont
  {Zhang}(2023)}]{jiao2023effect}%
  \BibitemOpen
  \bibfield  {author} {\bibinfo {author} {\bibfnamefont {L.}~\bibnamefont
  {Jiao}}\ and\ \bibinfo {author} {\bibfnamefont {X.}~\bibnamefont {Zhang}},\
  }\bibfield  {title} {\bibinfo {title} {Effect of earth-moon’s gravity on
  tianqin’s range acceleration noise. iii. an analytical model},\ }\href@noop
  {} {\bibfield  {journal} {\bibinfo  {journal} {Physical Review D}\ }\textbf
  {\bibinfo {volume} {107}},\ \bibinfo {pages} {102004} (\bibinfo {year}
  {2023})}\BibitemShut {NoStop}%
\bibitem [{\citenamefont {Jing}\ \emph {et~al.}(2022)\citenamefont {Jing},
  \citenamefont {Zheng}, \citenamefont {Yang}, \citenamefont {Zhang},
  \citenamefont {Lu}, \citenamefont {Tang},\ and\ \citenamefont
  {Su}}]{jing2022plasma}%
  \BibitemOpen
  \bibfield  {author} {\bibinfo {author} {\bibfnamefont {Y.-D.}\ \bibnamefont
  {Jing}}, \bibinfo {author} {\bibfnamefont {L.}~\bibnamefont {Zheng}},
  \bibinfo {author} {\bibfnamefont {S.}~\bibnamefont {Yang}}, \bibinfo {author}
  {\bibfnamefont {X.}~\bibnamefont {Zhang}}, \bibinfo {author} {\bibfnamefont
  {L.}~\bibnamefont {Lu}}, \bibinfo {author} {\bibfnamefont {B.}~\bibnamefont
  {Tang}},\ and\ \bibinfo {author} {\bibfnamefont {W.}~\bibnamefont {Su}},\
  }\bibfield  {title} {\bibinfo {title} {Plasma noise in tianqin time-delay
  interferometry},\ }\href@noop {} {\bibfield  {journal} {\bibinfo  {journal}
  {Physical Review D}\ }\textbf {\bibinfo {volume} {106}},\ \bibinfo {pages}
  {082006} (\bibinfo {year} {2022})}\BibitemShut {NoStop}%
\bibitem [{\citenamefont {Liu}\ \emph {et~al.}(2024)\citenamefont {Liu},
  \citenamefont {Su}, \citenamefont {Zhang}, \citenamefont {Zhang},\ and\
  \citenamefont {Zhou}}]{liu2024solar}%
  \BibitemOpen
  \bibfield  {author} {\bibinfo {author} {\bibfnamefont {Y.}~\bibnamefont
  {Liu}}, \bibinfo {author} {\bibfnamefont {W.}~\bibnamefont {Su}}, \bibinfo
  {author} {\bibfnamefont {X.}~\bibnamefont {Zhang}}, \bibinfo {author}
  {\bibfnamefont {J.}~\bibnamefont {Zhang}},\ and\ \bibinfo {author}
  {\bibfnamefont {S.}~\bibnamefont {Zhou}},\ }\bibfield  {title} {\bibinfo
  {title} {Solar plasma noise in tianqin laser propagation: An extreme case and
  statistical analysis},\ }\href@noop {} {\bibfield  {journal} {\bibinfo
  {journal} {The Astrophysical Journal}\ }\textbf {\bibinfo {volume} {975}},\
  \bibinfo {pages} {291} (\bibinfo {year} {2024})}\BibitemShut {NoStop}%
\bibitem [{\citenamefont {Ye}\ \emph {et~al.}(2023)\citenamefont {Ye},
  \citenamefont {Lian}, \citenamefont {Zhao},\ and\ \citenamefont
  {Zhang}}]{ye2023novel}%
  \BibitemOpen
  \bibfield  {author} {\bibinfo {author} {\bibfnamefont {X.}~\bibnamefont
  {Ye}}, \bibinfo {author} {\bibfnamefont {J.}~\bibnamefont {Lian}}, \bibinfo
  {author} {\bibfnamefont {G.}~\bibnamefont {Zhao}},\ and\ \bibinfo {author}
  {\bibfnamefont {D.}~\bibnamefont {Zhang}},\ }\bibfield  {title} {\bibinfo
  {title} {A novel closed-loop structure for drag-free control systems with
  eskf and lqr},\ }\href@noop {} {\bibfield  {journal} {\bibinfo  {journal}
  {Sensors}\ }\textbf {\bibinfo {volume} {23}},\ \bibinfo {pages} {6766}
  (\bibinfo {year} {2023})}\BibitemShut {NoStop}%
\bibitem [{\citenamefont {Zhang}\ \emph {et~al.}(2024)\citenamefont {Zhang},
  \citenamefont {Ye}, \citenamefont {Li}, \citenamefont {Zhao},\ and\
  \citenamefont {Lian}}]{zhang2024nonlinear}%
  \BibitemOpen
  \bibfield  {author} {\bibinfo {author} {\bibfnamefont {D.}~\bibnamefont
  {Zhang}}, \bibinfo {author} {\bibfnamefont {X.}~\bibnamefont {Ye}}, \bibinfo
  {author} {\bibfnamefont {H.}~\bibnamefont {Li}}, \bibinfo {author}
  {\bibfnamefont {G.}~\bibnamefont {Zhao}},\ and\ \bibinfo {author}
  {\bibfnamefont {J.}~\bibnamefont {Lian}},\ }\bibfield  {title} {\bibinfo
  {title} {Nonlinear modeling and validation of spacecraft dynamics for
  space-based gravitational wave detector},\ }\href@noop {} {\bibfield
  {journal} {\bibinfo  {journal} {Acta Astronautica}\ }\textbf {\bibinfo
  {volume} {224}},\ \bibinfo {pages} {57} (\bibinfo {year} {2024})}\BibitemShut
  {NoStop}%
\bibitem [{\citenamefont {Inchausp{\'e}}(2015)}]{inchauspe2015lisa}%
  \BibitemOpen
  \bibfield  {author} {\bibinfo {author} {\bibfnamefont {H.-R.}\ \bibnamefont
  {Inchausp{\'e}}},\ }\emph {\bibinfo {title} {De LISA Pathfinder {\`a} LISA:
  {\'E}laboration d’un simulateur dynamique pour la mission spatiale
  eLISA}},\ \href@noop {} {Ph.D. thesis},\ \bibinfo  {school} {Universit{\'e}
  Paris Diderot} (\bibinfo {year} {2015})\BibitemShut {NoStop}%
\bibitem [{\citenamefont {Prince}\ and\ \citenamefont
  {Dormand}(1981)}]{Prince1981}%
  \BibitemOpen
  \bibfield  {author} {\bibinfo {author} {\bibfnamefont {P.~J.}\ \bibnamefont
  {Prince}}\ and\ \bibinfo {author} {\bibfnamefont {J.~R.}\ \bibnamefont
  {Dormand}},\ }\bibfield  {title} {\bibinfo {title} {High order embedded
  {Runge-Kutta} formulae},\ }\href@noop {} {\bibfield  {journal} {\bibinfo
  {journal} {J. Comp. Appl. Math.}\ }\textbf {\bibinfo {volume} {7}},\ \bibinfo
  {pages} {67} (\bibinfo {year} {1981})}\BibitemShut {NoStop}%
\bibitem [{\citenamefont {Chauvineau}\ \emph {et~al.}(2005)\citenamefont
  {Chauvineau}, \citenamefont {Regimbau}, \citenamefont {Vinet},\ and\
  \citenamefont {Pireaux}}]{chauvineau2005relativistic}%
  \BibitemOpen
  \bibfield  {author} {\bibinfo {author} {\bibfnamefont {B.}~\bibnamefont
  {Chauvineau}}, \bibinfo {author} {\bibfnamefont {T.}~\bibnamefont
  {Regimbau}}, \bibinfo {author} {\bibfnamefont {J.-Y.}\ \bibnamefont
  {Vinet}},\ and\ \bibinfo {author} {\bibfnamefont {S.}~\bibnamefont
  {Pireaux}},\ }\bibfield  {title} {\bibinfo {title} {Relativistic analysis of
  the lisa long range optical links},\ }\href@noop {} {\bibfield  {journal}
  {\bibinfo  {journal} {Physical Review D—Particles, Fields, Gravitation, and
  Cosmology}\ }\textbf {\bibinfo {volume} {72}},\ \bibinfo {pages} {122003}
  (\bibinfo {year} {2005})}\BibitemShut {NoStop}%
\bibitem [{\citenamefont {Otto}(2015)}]{otto2015time}%
  \BibitemOpen
  \bibfield  {author} {\bibinfo {author} {\bibfnamefont {M.}~\bibnamefont
  {Otto}},\ }\emph {\bibinfo {title} {Time-delay interferometry simulations for
  the laser interferometer space antenna}},\ \href@noop {} {Ph.D. thesis},\
  \bibinfo  {school} {Hannover: Gottfried Wilhelm Leibniz Universit{\"a}t
  Hannover} (\bibinfo {year} {2015})\BibitemShut {NoStop}%
\bibitem [{\citenamefont {Han}\ \emph {et~al.}(2018)\citenamefont {Han},
  \citenamefont {Yong}, \citenamefont {Xu}, \citenamefont {Wang}, \citenamefont
  {Yang}, \citenamefont {Xue}, \citenamefont {Cai}, \citenamefont {Ren},
  \citenamefont {Peng},\ and\ \citenamefont {Pan}}]{han2018point}%
  \BibitemOpen
  \bibfield  {author} {\bibinfo {author} {\bibfnamefont {X.}~\bibnamefont
  {Han}}, \bibinfo {author} {\bibfnamefont {H.-L.}\ \bibnamefont {Yong}},
  \bibinfo {author} {\bibfnamefont {P.}~\bibnamefont {Xu}}, \bibinfo {author}
  {\bibfnamefont {W.-Y.}\ \bibnamefont {Wang}}, \bibinfo {author}
  {\bibfnamefont {K.-X.}\ \bibnamefont {Yang}}, \bibinfo {author}
  {\bibfnamefont {H.-J.}\ \bibnamefont {Xue}}, \bibinfo {author} {\bibfnamefont
  {W.-Q.}\ \bibnamefont {Cai}}, \bibinfo {author} {\bibfnamefont {J.-G.}\
  \bibnamefont {Ren}}, \bibinfo {author} {\bibfnamefont {C.-Z.}\ \bibnamefont
  {Peng}},\ and\ \bibinfo {author} {\bibfnamefont {J.-W.}\ \bibnamefont
  {Pan}},\ }\bibfield  {title} {\bibinfo {title} {Point-ahead demonstration of
  a transmitting antenna for satellite quantum communication},\ }\href@noop {}
  {\bibfield  {journal} {\bibinfo  {journal} {Optics express}\ }\textbf
  {\bibinfo {volume} {26}},\ \bibinfo {pages} {17044} (\bibinfo {year}
  {2018})}\BibitemShut {NoStop}%
\bibitem [{\citenamefont {Ye}\ \emph {et~al.}(2019)\citenamefont {Ye},
  \citenamefont {Zhang}, \citenamefont {Zhou}, \citenamefont {Wang},
  \citenamefont {Yuan}, \citenamefont {Gu}, \citenamefont {Ding}, \citenamefont
  {Zhang}, \citenamefont {Mei},\ and\ \citenamefont {Luo}}]{Ye2019}%
  \BibitemOpen
  \bibfield  {author} {\bibinfo {author} {\bibfnamefont {B.}~\bibnamefont
  {Ye}}, \bibinfo {author} {\bibfnamefont {X.}~\bibnamefont {Zhang}}, \bibinfo
  {author} {\bibfnamefont {M.}~\bibnamefont {Zhou}}, \bibinfo {author}
  {\bibfnamefont {Y.}~\bibnamefont {Wang}}, \bibinfo {author} {\bibfnamefont
  {H.}~\bibnamefont {Yuan}}, \bibinfo {author} {\bibfnamefont {D.}~\bibnamefont
  {Gu}}, \bibinfo {author} {\bibfnamefont {Y.}~\bibnamefont {Ding}}, \bibinfo
  {author} {\bibfnamefont {J.}~\bibnamefont {Zhang}}, \bibinfo {author}
  {\bibfnamefont {J.}~\bibnamefont {Mei}},\ and\ \bibinfo {author}
  {\bibfnamefont {J.}~\bibnamefont {Luo}},\ }\bibfield  {title} {\bibinfo
  {title} {Optimizing orbits for {TianQin}},\ }\href@noop {} {\bibfield
  {journal} {\bibinfo  {journal} {Int. J. Mod. Phys. D}\ }\textbf {\bibinfo
  {volume} {28}},\ \bibinfo {pages} {1950121} (\bibinfo {year}
  {2019})}\BibitemShut {NoStop}%
\bibitem [{\citenamefont {Chen}\ \emph {et~al.}(2023)\citenamefont {Chen},
  \citenamefont {Tan}, \citenamefont {Li}, \citenamefont {Zhu}, \citenamefont
  {Zhao}, \citenamefont {Zhang},\ and\ \citenamefont
  {Yang}}]{chen23selfgravity}%
  \BibitemOpen
  \bibfield  {author} {\bibinfo {author} {\bibfnamefont {B.}~\bibnamefont
  {Chen}}, \bibinfo {author} {\bibfnamefont {W.}~\bibnamefont {Tan}}, \bibinfo
  {author} {\bibfnamefont {W.}~\bibnamefont {Li}}, \bibinfo {author}
  {\bibfnamefont {L.}~\bibnamefont {Zhu}}, \bibinfo {author} {\bibfnamefont
  {H.}~\bibnamefont {Zhao}}, \bibinfo {author} {\bibfnamefont {X.}~\bibnamefont
  {Zhang}},\ and\ \bibinfo {author} {\bibfnamefont {S.}~\bibnamefont {Yang}},\
  }\href@noop {} {\emph {\bibinfo {title} {Self-gravity Analysis for TianQin
  Satellite (in chinese)}}},\ \bibinfo {type} {Tech. Rep.}\ (\bibinfo
  {institution} {TianQin Research Center for Gravitational Physics},\ \bibinfo
  {year} {2023})\BibitemShut {NoStop}%
\bibitem [{\citenamefont {Wang}\ \emph {et~al.}(2024)\citenamefont {Wang},
  \citenamefont {Zhang},\ and\ \citenamefont {Duan}}]{wang2024point}%
  \BibitemOpen
  \bibfield  {author} {\bibinfo {author} {\bibfnamefont {D.}~\bibnamefont
  {Wang}}, \bibinfo {author} {\bibfnamefont {X.}~\bibnamefont {Zhang}},\ and\
  \bibinfo {author} {\bibfnamefont {H.-Z.}\ \bibnamefont {Duan}},\ }\bibfield
  {title} {\bibinfo {title} {On point-ahead angle control strategies for
  tianqin},\ }\href@noop {} {\bibfield  {journal} {\bibinfo  {journal}
  {Classical and Quantum Gravity}\ }\textbf {\bibinfo {volume} {41}},\ \bibinfo
  {pages} {117003} (\bibinfo {year} {2024})}\BibitemShut {NoStop}%
\bibitem [{\citenamefont {Zheng}\ \emph {et~al.}(2023)\citenamefont {Zheng},
  \citenamefont {Yang},\ and\ \citenamefont {Zhang}}]{zheng2023doppler}%
  \BibitemOpen
  \bibfield  {author} {\bibinfo {author} {\bibfnamefont {L.}~\bibnamefont
  {Zheng}}, \bibinfo {author} {\bibfnamefont {S.}~\bibnamefont {Yang}},\ and\
  \bibinfo {author} {\bibfnamefont {X.}~\bibnamefont {Zhang}},\ }\bibfield
  {title} {\bibinfo {title} {Doppler effect in tianqin time-delay
  interferometry},\ }\href@noop {} {\bibfield  {journal} {\bibinfo  {journal}
  {Physical Review D}\ }\textbf {\bibinfo {volume} {108}},\ \bibinfo {pages}
  {022001} (\bibinfo {year} {2023})}\BibitemShut {NoStop}%
\bibitem [{\citenamefont {Vidano}\ \emph {et~al.}(2020)\citenamefont {Vidano},
  \citenamefont {Novara}, \citenamefont {Colangelo},\ and\ \citenamefont
  {Grzymisch}}]{vidano2020lisa}%
  \BibitemOpen
  \bibfield  {author} {\bibinfo {author} {\bibfnamefont {S.}~\bibnamefont
  {Vidano}}, \bibinfo {author} {\bibfnamefont {C.}~\bibnamefont {Novara}},
  \bibinfo {author} {\bibfnamefont {L.}~\bibnamefont {Colangelo}},\ and\
  \bibinfo {author} {\bibfnamefont {J.}~\bibnamefont {Grzymisch}},\ }\bibfield
  {title} {\bibinfo {title} {The lisa dfacs: A nonlinear model for the
  spacecraft dynamics},\ }\href@noop {} {\bibfield  {journal} {\bibinfo
  {journal} {Aerospace Science and Technology}\ }\textbf {\bibinfo {volume}
  {107}},\ \bibinfo {pages} {106313} (\bibinfo {year} {2020})}\BibitemShut
  {NoStop}%
\end{thebibliography}%
\end{document}